\documentclass[fleqn,usenatbib]{mnras}

\usepackage{newtxtext,newtxmath}

\usepackage[T1]{fontenc}

\DeclareRobustCommand{\VAN}[3]{#2}
\let\VANthebibliography\thebibliography
\def\thebibliography{\DeclareRobustCommand{\VAN}[3]{##3}\VANthebibliography}

\usepackage{graphicx}	
\usepackage{amsmath}	
\usepackage[flushleft]{threeparttable}
\usepackage{multirow}
\usepackage{array}
\usepackage{xcolor}

\newcommand{\ps}{\,s$^{-1}$}
\newcommand{\flux}{\,erg\,s$^{-1}$\,cm$^{-2}$}
\newcommand{\NH}{$N_{\rm H}$}
\newcommand{\LxLb}{$L_\text{X}/L_\text{bol}$}

\title[XUV irradiation of Praesepe planets]{The strongly irradiated planets in Praesepe}

\author[G. W. King et al.]{George W. King,$^{1,2,3}$\thanks{E-mail: kinggw@umich.edu}
Peter J. Wheatley,$^{2,3}$
Victoria A. Fawcett,$^{4,2}$
Nicola J. Miller,$^{5,2}$
L\'{i}a R. Corrales,$^1$
\newauthor and Marcel A.~Ag\"{u}eros$^6$
\\
$^{1}$Department of Astronomy, University of Michigan, Ann Arbor, MI 48109, USA\\
$^{2}$Department of Physics, University of Warwick, Gibbet Hill Road, Coventry, CV4 7AL, UK\\
$^{3}$Centre for Exoplanets and Habitability, University of Warwick, Gibbet Hill Road, Coventry, CV4 7AL, UK\\
$^{4}$Centre for Extragalactic Astronomy, Department of Physics, Durham University, DH1 3LE, UK\\
$^{5}$Astrophysics Group, Keele University, Staffordshire, ST5 5BG, UK\\
$^{6}$Department of Astronomy, Columbia University, 550 West 120th Street, New York, NY 10027, USA
}

\date{Accepted XXX. Received YYY; in original form ZZZ}

\pubyear{2021}

\begin{document}
\label{firstpage}
\pagerange{\pageref{firstpage}--\pageref{lastpage}}
\maketitle

\begin{abstract}
We present an analysis of \textit{XMM-Newton} observations of four stars in the young (670\,Myr) open cluster Praesepe. The planets hosted by these stars all lie close in radius-period space to the radius-period valley and/or the Neptunian desert, two features that photoevaporation by X-ray and extreme ultraviolet (EUV) photons could be driving. Although the stars are no longer in the saturated regime, strong X-ray and extreme ultraviolet irradiation is still ongoing. Based on EUV time evolution slopes we derived in a previous paper, in all four cases, two-thirds of their EUV irradiation is still to come. We compare the \textit{XMM-Newton} light curves to those simultaneously measured with \textit{K2} at optical wavelengths, allowing us to search for correlated variability between the X-ray and optical light curves. We find that the X-ray flux decreases and flattens off while the optical flux rises throughout for K2-100, something that could result from active regions disappearing from view as the star spins. Finally, we also investigate possible futures for the four planets in our sample with simulations of their atmosphere evolution still to come, finding that complete photoevaporative stripping of the envelope of three of the four planets is possible, depending on the current planet masses.
\end{abstract}




\section{Introduction}
The observed distribution of exoplanets in radius-orbital period space includes two regions with a deficit of planets that cannot be explained by selection biases. The first of these, the ``radius-period valley,'' is an observed dearth of planets with radii between 1.5 and 2.0\,R$_\oplus$, uncovered in the \textit{Kepler} planets \citep{Fulton2017,Fulton2018}. Further analysis of a smaller sample by~\citet{VanEylen2018}, using asteroseismology for the derivation of the stellar properties, showed evidence of both a negative slope in radius-period space, which \citet{Martinez2019} revealed further evidence of, yielding a consistent value of the gradient. The \citet{VanEylen2018} study also inferred the valley to be a clean break with few or no planets existing in the gap, suggestive of a homogeneous core composition across their sample. If this is the case, then the exact position of the gap in radius and the gradient of the slope in radius-period space is indicative of the composition of the planets, with observations pointing to a rocky composition \citep[][hereafter OW17; \citealt{Jin2018}]{Owen2017}. \citet{Venturini2020} have reproduced the bimodality with two different core compositions, although combinations of different compositions (e.g. a 50-50 ratio of rock and ice) might be expected to populate the gap, something which has not been observed. 

This radius-period valley had been predicted~\citep{Owen2013,Lopez2013,Jin2014,Chen2016}, and there are two proposed mechanisms that can reproduce the observed valley: photoevaporation and core-powered mass loss. In the former, atmospheric photoevaporation, driven by X-ray and extreme ultraviolet (EUV; together, XUV) photons, strips planets born with small H/He envelopes down to bare rocky cores, with at most a secondary atmosphere of heavier elements (e.g. OW17). The gap's position and slope are consistent with a photoevaporation origin, as well as a population whose core compositions are Earth-like \citep[e.g. OW17;][]{Jin2018,Martinez2019}. In core-powered mass loss the atmospheric escape is driven by the gradual dissipation of the energy in the planetary core following its formation, and it too has been found to produce a valley with a position and slope that match observations \citep{Ginzburg2016,Gupta2019,Gupta2020}. \citet{Rogers2021} recently discussed possible methods of distinguishing which mechanism, if either, dominates, but for now conclusions are uncertain. For the purposes of this work, we focus only on the effects of photoevaporation in the systems investigated.

The other region in radius-period space lacking in observed exoplanets is the ``Neptunian desert,'' which is an observed dearth of intermediate-sized planets at short orbital periods \citep{Szabo2011,Beauge2013,Helled2016,Lundkvist2016,Mazeh2016,Owen2018}. This effect, also observed in the mass-period plane, has also been attributed to photoevaporation~\citep{Kurokawa2014}.~\citet{Mazeh2016} showed the region to be triangular-shaped (see Fig.~\ref{fig:RpPorb}), and derived empirical relations for the boundaries. Possible evaporation-forbidden regions of the parameter space had been discussed~\citep[e.g.][]{LDE2007}. \citet{Owen2018} used numerical models to show that photoevaporation can explain the lower boundary. These authors also suggest the upper boundary may extend too high up in both planes of interest for those processes to be the sole origin of the desert, and that a better match to the boundary may be the tidal disruption barrier for planets migrating inwards by high-eccentricity excitation. This latter process had previously been discussed in the context of the Neptunian desert by~\citet{Matsakos2016}. Again, we solely consider photoevaporative processes in this study.

Since stellar emission of X-ray photons is at its highest in the first few hundred Myr of a star's life~\citep[e.g.][]{Micela1985,Guedel1997,Micela2002,Feigelson2004,Jackson2012}, it is at these early times that these features are thought to be carved out if photoevaporation is indeed the driving mechanism. This assumption partially relies on the EUV and X-ray irradiation falling off at a high enough rate that the total integrated XUV irradiation at late times does not exceed that for the first few hundred Myr. However, we recently explored the time evolution of EUV emission in \citet{EUVevolution}, finding that EUV emission declines much less quickly than X-rays. This could mean that the timescale for significant photoevaporation is longer than the first 100\,Myr, which would make the observational signature of its effects on the population more difficult to separate from core-powered mass loss, whose timescale is thought to be on the order of a Gyr \citep{Gupta2019}.

A big observational hindrance to exploring this further is the relatively few known planets younger than a Gyr. The vast majority of currently known exoplanets orbit mature stars, with typical ages of a few Gyr. This is primarily because the stellar activity of young stars poses challenges for the discovery and characterisation of exoplanets, particularly via the radial velocity method. Furthermore, the target stars of the prolific \textit{Kepler} mission were typically old field stars, as the field-of-view was deliberately chosen to avoid young stellar populations~\citep{Batalha2010}. The repurposed  \textit{Kepler} mission, known as \textit{K2} \citep{Howell2014}, targeted a number of open clusters spanning a range of young and intermediate ages searching for planets. However, with comparatively few such planets still discovered, the known population remains far too small for statistically meaningful comparisons of the exoplanet properties (e.g. in radius-period space) for systems of different ages. This means the epoch at which features such as the radius-period valley or the Neptunian desert are forged cannot be determined in this way for the foreseeable future, even with the exoplanet discoveries coming out of the TESS mission including some young planets \citep[e.g.][]{Newton2019,Tofflemire2021}.

Among the discoveries \textit{K2} made for systems in open clusters is that of nine planetary candidates orbiting eight stars in Praesepe (also known as M44, or the Beehive Cluster). In Table~\ref{tab:K2Praesepe}, we summarise these eight systems, together with providing references for their discoveries in the literature. In this work, we study the high-energy environments of the planets in the K2-95, K2-100, K2-101, and K2-104 systems. These are all planets that were validated in \citet[][hereafter M17]{Mann2017} with false positive probabilities of less than 1 per cent. Among our sample, \citet{Barragan2019} have since successfully measured a radial velocity signal for K2-100, providing the first ever mass measurement of a planet in a young open cluster.

\begin{table}
  \caption{Planets discovered by in Praesepe by the \textit{K2} mission, together with literature references.}
  \label{tab:K2Praesepe}
  \centering
  \begin{threeparttable}
\begin{tabular}{ccc}
\hline
System         & \begin{tabular}[c]{@{}c@{}}Number of\\ planets\end{tabular} & References \\
\hline
K2-95          & 1                                                           & 1, 2, 5 \\
K2-100         & 1                                                           & 1, 3, 4, 5 \\
K2-101         & 1                                                           & 1, 3, 5 \\
K2-102         & 1                                                           & 5 \\
K2-103         & 1                                                           & 1, 3, 5 \\
K2-104         & 1                                                           & 4, 5 \\
K2-264         & 2                                                           & 6, 7 \\
EPIC 211901114 & 1*                                                          & 5 \\
\hline
\end{tabular}
    \begin{tablenotes}
\item $^*$ Planet is an unconfirmed candidate.
\item References: (1) \citet{Libralato2016}; (2) \citet{Obermeier2016}; (3) \citet{Barros2016}; (4) \citet{Pope2016}; (5) M17; (6) \citet{Livingston2018}; (7) \citet{Rizzuto2018}.
    \end{tablenotes}
  \end{threeparttable}
\end{table}

As members of a cluster, these systems have a more reliable age estimate than most known planetary systems, which tend to orbit field stars where ages are harder to constrain. Indeed, cluster ages are often used to calibrate age determination methods for use on field stars~\citep[for a review, see][]{Soderblom2010}. At an intermediate age of $670\pm67$ Myr years old~\citep{Douglas2019}, these planets in Praesepe are also likely more intensely irradiated at the current epoch than most systems previously studied at high energies, and so perhaps still undergoing significant atmospheric mass loss. Moreover, the estimated radii for most of these planets is consistent with being in or close to the radius-period valley and/or Neptunian desert, making them an ideal testing ground for these theories of atmosphere evolution.

\section{Our Sample}
\label{sec:sample}

\begin{figure}
\centering
 \includegraphics[width=0.90\columnwidth]{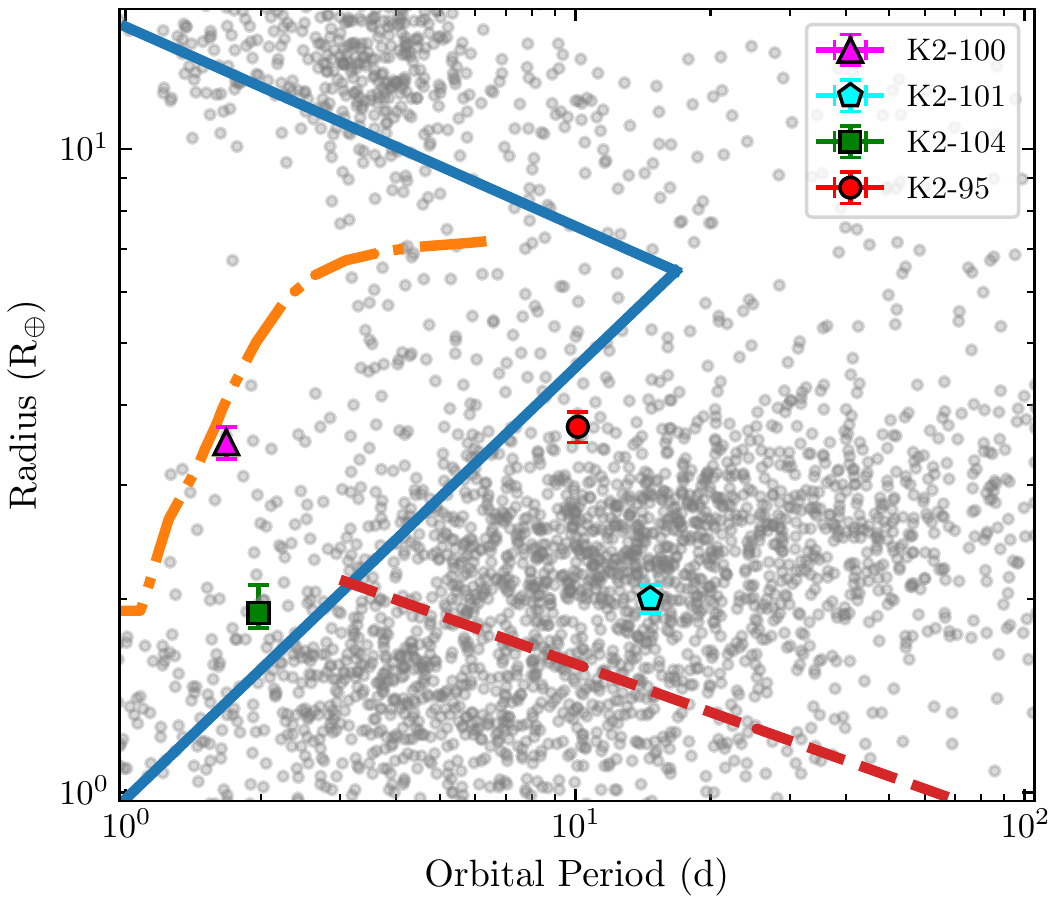}
 \caption{Positions of the four planets in our sample in radius-period space. The solid blue triangular region is the Neptunian desert, as empirically determined by~\citet{Mazeh2016}. The dashed red line corresponds to the most massive planet than can have been stripped bare at a given period, as analytically determined by OW17 (and thus provides an estimate of the position of the radius-period valley). The dot-dash orange line is one of the numerical solutions for the lower desert boundary determined by OW17 for a core mass of 13.75\,M$_\oplus$, the largest core mass they plotted. The grey points are the full population of exoplanets with a measured radius and period.} 
 \label{fig:RpPorb}
\end{figure}

\begin{table*}
  \caption[Praesepe sample system parameters]{Parameters for the systems in our sample.}
  \label{tab:SysParam}
  \centering
  \begin{threeparttable}
\begin{tabular}{lccccccccccccc}
\hline
System	& Spectral & $V$	& $d^a$ & $R_*$	 & $T_{\rm eff,*}$& $L_{\rm bol}$& $P_{\rm rot}$ & $R_{\rm p}$	& $P_{\rm orb}^b$& $a$	  & $T_0^b$     & $T_{\rm eq}^c$& Ref.$^d$\\
	& Type     & (mag)	& (pc)  & ($R_\odot$) & (K)		  & (L$_\odot$)  &(d)		 & ($R_\oplus$) & (d)		 & (au)   &($\dagger$)& (K)\\
		\hline
K2-95	& M2V	   & 17.22$^e$	& 180.1 & 0.44	 & 3410		  & 0.0232       & 23.9		 & 3.7		& 10.135091	 & 0.0679 & 140.74083 & 419		& M17\\
K2-100	& G0V	   & 10.52$^f$	& 182.7 & 1.24	 & 5945		  & 1.72         & 4.3		 & 3.88		& 1.6739035	 & 0.0301 & 140.71941 & 1841		& \citet{Barragan2019}\\
K2-101	& K3V	   & 12.96$^f$	& 185.3 & 0.73	 & 4819		  & 0.2542       & 10.6		 & 2.0		& 14.677286	 & 0.1103 & 152.68135 & 598		& M17\\
K2-104	& M1V	   & 16.36$^e$	& 187.2 & 0.48	 & 3660		  & 0.0368       & 9.3		 & 1.9		& 1.974190	 & 0.2455 & 140.38117 & 780		& M17\\
\hline
\end{tabular}
    \begin{tablenotes}
\item $^a$ Converted from \textit{Gaia} EDR3 parallaxes.
\item $^b$ Ephemerides are BJD$_{\rm TDB}$ - 2457000.
\item $^c$ Equilibrium temperature calculated assuming zero Bond albedo and uniform redistribution of heat.
\item $^d$ Source of parameters and ephemerides, except where otherwise indicated.
\item $^e$ Converted from SDSS g and r \citep{Wang2014}, using~\citet{Jordi2006}.
\item $^f$ Measured with APASS~\citep{Henden2012}.
    \end{tablenotes}
  \end{threeparttable}
\end{table*}

Our sample consists of K2-95, K2-100, K2-101, and K2-104. The planets and their host stars are detailed in Table~\ref{tab:SysParam}, where the adopted values are from \citet{Barragan2019} for K2-100 and M17 for K2-95, K2-101 and K2-104, unless stated otherwise. One exception is for the distances, for which we calculate individual distances for each system using the parallaxes from the Gaia Early Data Release 3~\citep[EDR3;][]{GaiaEDR3}. All four planets reside close to the Neptunian desert, radius-period valley, or both. They are therefore important in the context of studying these features in the exoplanet population, particularly given their relative youth and potential ongoing evolution.

The four planets are plotted in radius-period space in Fig.~\ref{fig:RpPorb}. Also plotted are lines showing the empirical Neptunian desert~\citep{Mazeh2016}, a numerical determination of the lower desert boundary~\citep[for a core mass of 13.75\,M$_\oplus$]{Owen2018}, and an analytical determination of the largest planet that can be a stripped core (OW17). Note also that for K2-101's orbital period, the approximate location of the upper boundary of the radius valley is 2\,R$_\oplus$, similar to the measured radius \citep[e.g.][]{VanEylen2018,Martinez2019}. We follow OW17 in only plotting the latter for periods greater than 3\,d. OW17 also note that the evaporation and core composition models used affect the exact scaling and vertical position of the line, respectively. Similarly, the lower desert boundary from~\citet{Owen2018} plotted is one of a set of solutions the study finds, depending on the core mass and atmospheric metallicity. We refer the reader to their paper and figures within for the full range of boundary solutions they determine, where the other lines are all below the representative line plotted here. 

Fig.~\ref{fig:RpPorb} highlights the proximity of the planets to features likely to be associated with evaporation in the exoplanet population. K2-100b is seen to lie well inside the \citet{Mazeh2016} desert. It is also inside the desert for most, though not all, solutions from~\citet{Owen2018}. K2-104b also resides within the \citet{Mazeh2016} desert. However, it lies below all the~\citet{Owen2018} solutions for the desert, and if one extrapolates the lower boundary of the evaporation valley to shorter periods it would lie below this line too, suggesting it could actually be a stripped core of a once larger planet. K2-101b intriguingly resides just a bit above the radius-period valley, and so could either retain a small envelope for the rest of its life, or could still be in process of being stripped. K2-95b's position is close to the \citet{Mazeh2016} desert and so maybe it evolved out of this region, although the~\citet{Owen2018} solutions place it a bit further away from the desert boundary.

\section{Observations}
\label{sec:Obs}
\subsection{\textit{XMM-Newton}}
\label{ssec:XMMobs}

\begin{table*}
  \caption[Praesepe sample \textit{XMM-Newton} observations]{Details of the \textit{XMM-Newton} observations.}
  \label{tab:Obs}
  \centering
  \begin{threeparttable}
  \begin{tabular}{c c c c l c c c c}
    \hline\
    ObsID & PI & Start time		& Exp. T & System(s) & Start -- Stop	& Transit		& EPIC			& EPIC	  \\
    		       &    & (TDB)\,$^a$	& (ks)\,$^a$ &	& phase\,$^b$	& phase\,$^b$	& Cameras\,$^c$	& filter  \\
    \hline
    \multirow{2}{*}{0721620101} & \multirow{2}{*}{Ag\"ueros} & \multirow{2}{*}{2013-10-30 11:34} & \multirow{2}{*}{68.8} & K2-100 & 0.573 -- 1.057 & 0.981 -- 1.019 & All & \multirow{2}{*}{Thin} \\
    			&		& 				   & 	  & K2-104 & 0.241 -- 0.652 & 0.987 -- 1.013 & pn, MOS2	& \\[0.1cm]
    0761920901 & Drake & 2015-05-06 00:36 & 59.9 & K2-100 & 0.662 -- 1.077 & 0.981 -- 1.057 & All		& Medium \\
    0761921001 & Drake & 2015-05-09 13:13 & 50.0 & K2-101 & 0.957 -- 0.998 & 0.995 -- 1.005 & pn, MOS2	& Medium \\
    0761921101 & Drake & 2015-05-11 23:19 & 61.2 & K2-95  & 0.355 -- 0.426 & 0.995 -- 1.005 & All		& Medium \\
    \hline
\end{tabular}
    \begin{tablenotes}
\item $^a$ Start time and duration are given for EPIC-pn.
\item $^b$ All ephemerides taken from~\citet{Mann2017}, except for K2-100, which are from \citet{Barragan2019}.
\item $^c$ Listing of which of the EPIC cameras' fields of view the source appears in. The three EPIC cameras are pn, MOS1, and MOS2.
    \end{tablenotes}
  \end{threeparttable}
\end{table*}

We have analysed archival \textit{XMM-Newton} \citep{Jansen2001} observations that together include data for the four systems introduced in Section~\ref{sec:sample}. These observations are summarised in Table~\ref{tab:Obs}. K2-101, K2-95, and K2-104 have each been observed once, while K2-100 has been observed twice: once in 2013 when K2-104 was also within the field of view, and a second time in 2015. While the \textit{XMM-Newton} Science Archive\footnote{\url{http://nxsa.esac.esa.int}} lists an additional observation for K2-102 (ObsID: 0101440401; PI: Pallavicini), inspection of the images revealed the source to have fallen just outside the field of view of all three EPIC cameras. 
Further, two of the objects, K2-101 and K2-104, fell on to one of the faulty CCDs of the EPIC-MOS1 camera during their respective observations. These two stars therefore only have data from EPIC-pn and EPIC-MOS2.

Only K2-95 was observed by the Optical Monitor (OM). None of the observations specifically targeted these systems, and so the other objects fell outside the far smaller field of view of the OM. The K2-95 OM data were taken with the UVM2 filter (effective wavelength 2310\,\AA; width 480\,\AA).

The data were reduced using the Scientific Analysis System (\textsc{sas} 16.0.0). The standard procedures were used in each case\footnote{As outlined on the `SAS Threads' webpages: \url{http://www.cosmos.esa.int/web/xmm-newton/sas-threads}}. Elevated high-energy background, due to Solar soft protons~\citep{Walsh2014}, affected a small proportion of each observation. These periods were filtered out during the spectral fitting and ensuing analyses. However, the unfiltered data were used to produce the light curves in Section~\ref{ssec:xlc}, in order to avoid  gaps.

\subsection{K2 light curves}
Praesepe was observed in 2015 between April 27 and July 10 as part of campaign 5 of the \textit{K2} mission \citep{Howell2014}. Three of the \textit{XMM-Newton} observations described above were simultaneous with the \textit{K2} observations: the 2015 observations of K2-100, K2-101, and K2-95. Praesepe was later observed again by \textit{K2} in campaigns 16 and 18, but neither of these were simultaneous with any \textit{XMM-Newton} observations, and so we do not consider them any further in this work.

Using the campaign 5 observations, we analysed the spot modulation of each star in its K2 light curve, in order to determine the modulation and phase at the time of the \textit{XMM-Newton} observations. We obtained the detrended, corrected \textit{K2} light curves output by the \textsc{everest} pipeline~\citep{Luger2016,Luger2017}. These were downloaded from the Barbara A. Mikulski Archive for Space Telescopes (MAST).

\section{\textit{XMM-Newton} results}
\label{sec:XMM}
K2-100 and K2-101 were both very clearly detected at the expected position in all three EPIC cameras. Restricting the MOS2 images for K2-95 to 0.6--1.0\,keV, energies expected to be bright for young objects due to Fe L-shell emission, showed an excess of counts at the expected position of the star. Using the same energy band in the pn for K2-104 revealed a marginal detection.

In analysing each observation, we employed 15\,arcsec radius source regions, centred on the proper-motion-corrected positions of each system. Multiple background regions from the same CCD chip were used for background subtraction. K2-104 is close on the sky to a galaxy cluster, the outskirts of which contribute additional background to the source region. We attempted to mitigate this by placing background regions on an arc about the centre of the galaxy cluster on the chip, such that the contamination would be similar in each background region to that in the source region.

The only target falling in the OM field of view, K2-95, was not detected in those observations using the UVM2 filter. The source detection algorithm applied by the standard reduction chain did not detect any source within 30\,arcsec of the expected position of the star. Visual inspection of the images confirmed the non-detection.

\subsection{X-ray light curves}
\label{ssec:xlc}

\begin{figure*}
\centering
 \includegraphics[width=\textwidth]{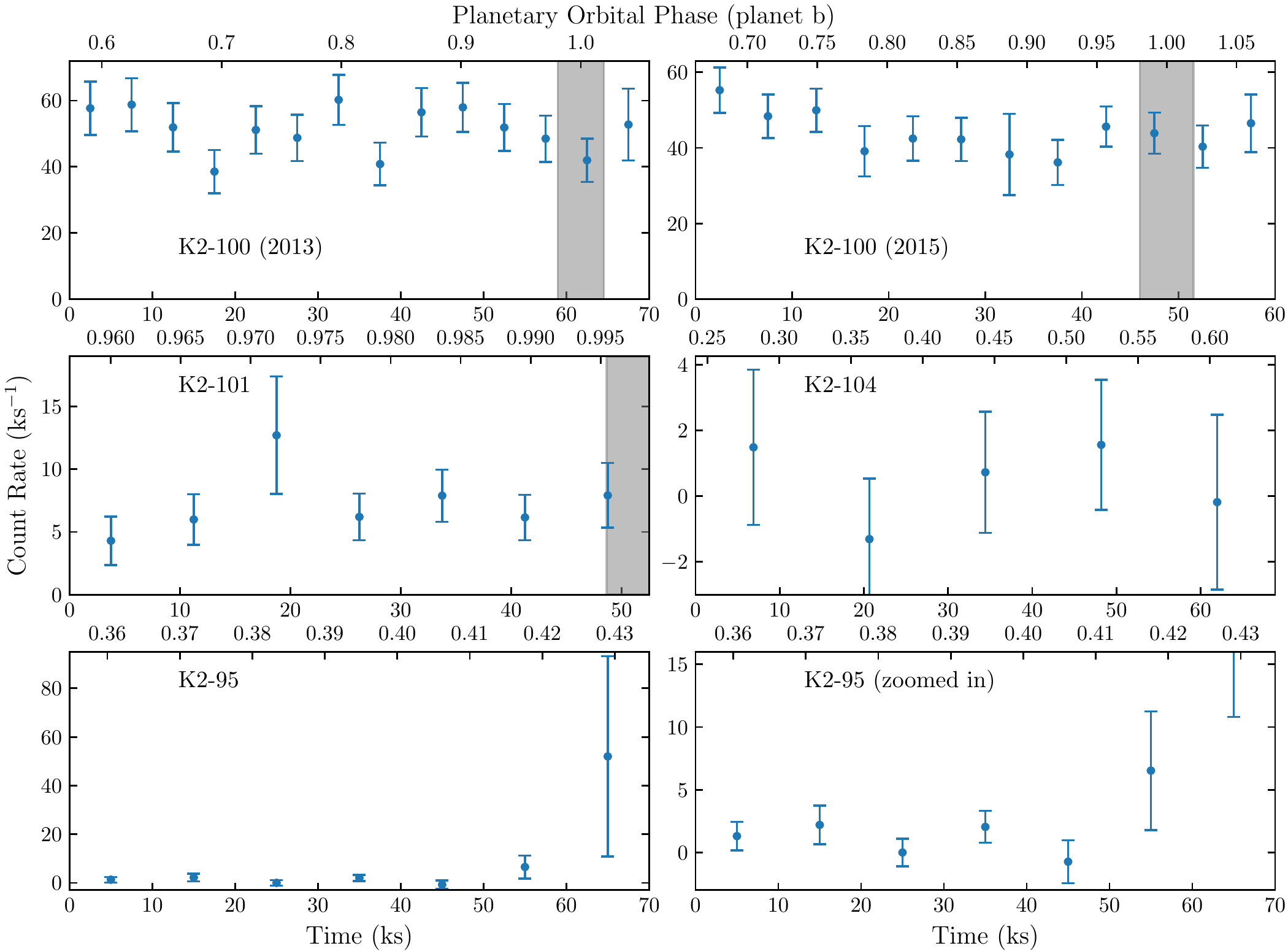}
 \caption{Background corrected X-ray light curves of the observations, covering the energy range 0.2--2.5\,keV. The count rate is the sum of all EPIC detectors for which data were available. Areas shaded in grey are the planetary transits (1st to 4th contact) in visible light. The bottom right panel for K2-95 is identical to the bottom left except for being zoomed in on the y-axis.}
 \label{fig:xlc}
\end{figure*}

We coadded the count rates across all EPIC cameras for which data were available for each of the observations. The resulting X-ray light curves are plotted in Fig.~\ref{fig:xlc}, covering the energy range 0.2--2.4\,keV. Transit phases in the optical are displayed as grey shaded regions. The light curves were searched for temporal variation in the X-ray flux, as well as for hints of transit features.

None of the observations show any strong flares, and none of the three light curves covering either a full or partial transit show any evidence of transit features. Both K2-100 datasets exhibit some variation in the count rate. The 2013 data dips down in a few places, while the 2015 data slopes down over the first 20\,ks before flattening off for the rest of the observation. This is explored further in the context of the simultaneous \textit{K2} data in Section~\ref{sec:K2}. The other three light curves have a low count rate, making them insensitive to low amplitude variability. The points with large error bars in each of the K2-100 (2015), K2-101, and K2-95 light curves correspond to periods of elevated background and are not indicative of any genuine source variability.

\subsection{X-ray spectra}
\label{ssec:XraySpec}

\begin{figure*}
\centering
 \includegraphics[width=0.94\textwidth]{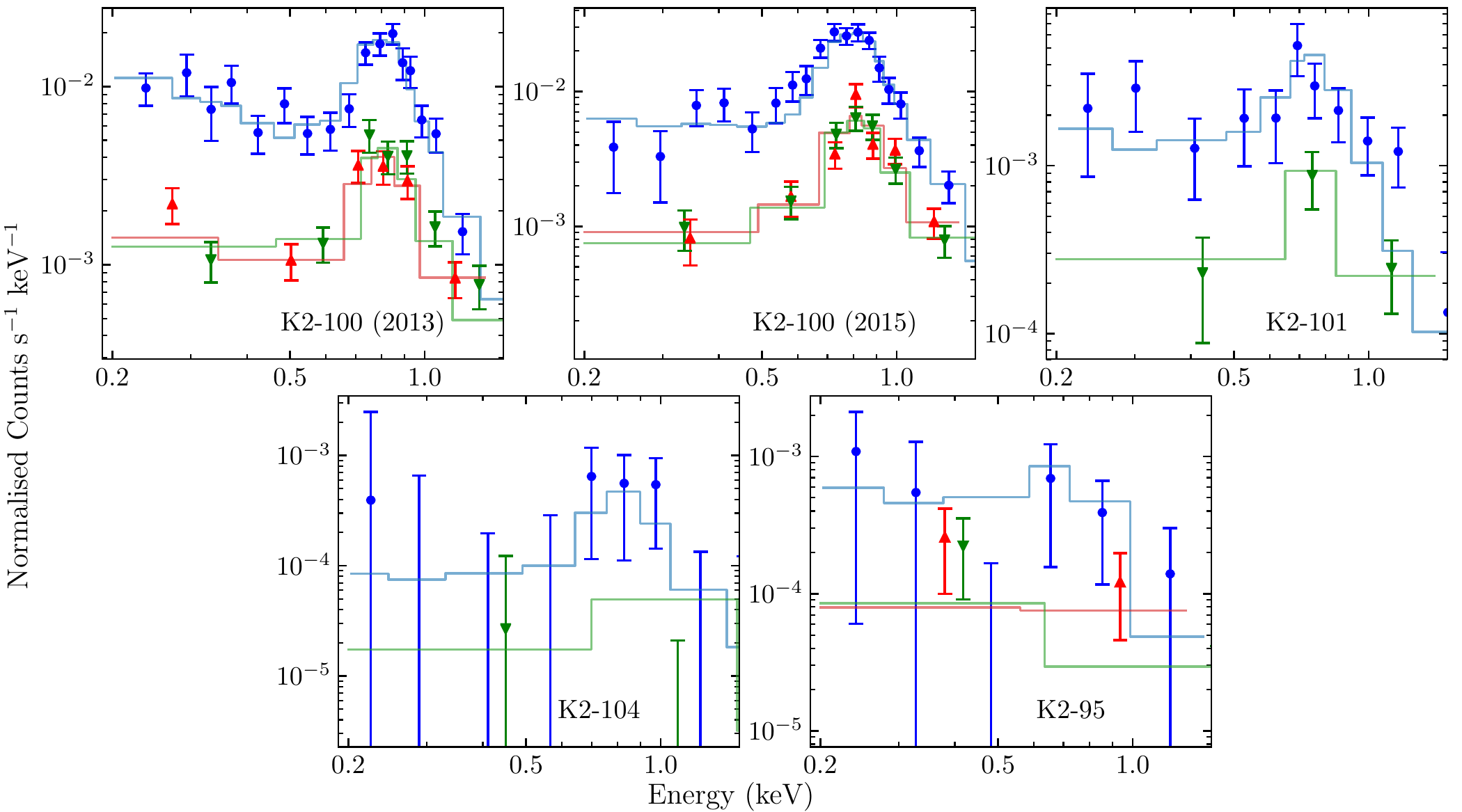}
 \caption[Praesepe EPIC spectra]{Observed X-ray spectra for each of the observations, displayed along with the best fit model. Each EPIC camera is displayed separately: EPIC-pn is shown with blue circles, EPIC-MOS1 with red up-pointing triangles, and EPIC-MOS2 with green down-pointing triangles.}
 \label{fig:spec}
\end{figure*}

\begin{table*}
\centering
\caption{Results of the X-ray spectral analysis. All fluxes and luminosities are for the 0.2--2.4\,keV band, except for the final two columns, which are for the 0.1--2.4\,keV `ROSAT' band. The latter all have `0.1' added in subscript to their column header. All fluxes are the unabsorbed flux.}
\label{tab:fit+fluxes}
\begin{threeparttable}
\begin{tabular}{lccccccccc}
\hline
System        & kT							& EM						& $F_{\rm X}$ 			& $L_{\rm X}$			& $L_{\rm EUV}$			& $F_{\rm XUV,\, p}$		& $F_{\rm XUV,\,1\,au}$ & $L_{\rm X,0.1}$	& $\dfrac{L_{\rm X,0.1}}{L_{\rm bol}}$ \\[0.2cm]
              & (keV)						& (\textit{a})				& (\textit{b})			& (\textit{c})			& (\textit{c})			& (\textit{d})				& (\textit{e})			& 	(\textit{c})	& ($\times10^{-5}$) \\
\hline
K2-100 (2013) & \begin{tabular}[c]{@{}c@{}}$0.0914^{+0.001}_{-0.031}$\\[0.05cm] $0.628^{+0.014}_{-0.026}$\end{tabular} & \begin{tabular}[c]{@{}c@{}}$114^{+424}_{-3}$\\[0.05cm] $63.5^{+4.4}_{-1.8}$\end{tabular} & $65.5^{+7.0}_{-1.1}$ & $262^{+28}_{-5}$ & $243^{+70}_{-51}$ & $187^{+27}_{-20}$ & $179^{+25}_{-18}$ & $434^{+85}_{-5}$ & $6.4^{+1.4}_{-0.8}$ \\[0.36cm]
K2-100 (2015) & \textit{As 2013}			& \begin{tabular}[c]{@{}c@{}}$25^{+68}_{-9}$\\[0.05cm] $63.8^{+3.4}_{-2.2}$\end{tabular} & $46.1^{+2.5}_{-2.1}$ & $184^{+10}_{-8}$ & $203^{+44}_{-42}$ & $144^{+17}_{-16}$ & $138^{+16}_{-15}$ & $238^{+18}_{-11}$ & $3.49^{+0.44}_{-0.45}$ \\[0.37cm]
K2-101        & $0.365^{+0.077}_{-0.034}$	& $15.0^{+1.7}_{-1.3}$		& $8.2\pm1.0$			& $33.5\pm4.3$			& $50\pm13$				& $2.44\pm0.40$				& $30.0\pm4.8$	& $37.3^\pm4.7$		& $3.67\pm0.49$ \\[0.2cm]
K2-104        & $0.71^{+0.52}_{-0.48}$		& $1.9^{+3.3}_{-0.8}$		& $1.2^{+1.0}_{-0.9}$	& $4.9^{+4.3}_{-3.9}$	& $13^{+12}_{-11}$		& $9.9^{+7.2}_{-6.5}$		& $6.2^{+4.6}_{-4.2}$	& $5.5^{+57.1}_{-3.5}$	& $3.7^{+38.1}_{-2.3}$ \\[0.2cm]
K2-95         & $0.27^{+0.80}_{-0.19}$		& $3.6^{+3.1}_{-1.7}$		& $2.01^{+0.74}_{-0.87}$	& $7.8^{+2.9}_{-3.4}$	& $14.7^{+7.2}_{-8.1}$	& $1.69^{+0.59}_{-0.62}$	& $8.0^{+2.7}_{-3.1}$	& $8.9^{+4.6}_{-3.8}$	& $10.1^{+5.3}_{-4.3}$ \\
\hline
\end{tabular}
    \begin{tablenotes}
\item \textit{a} $10^{50}$\,cm$^{-3}$ (Emission measure)
\item \textit{b} $10^{-15}$\,\flux\ (at Earth, unabsorbed) 
\item \textit{c} $10^{27}$\,erg\,s$^{-1}$
\item \textit{d} $10^{3}$\,\flux\
\item \textit{e} \flux\
    \end{tablenotes}
  \end{threeparttable}
\end{table*}

The X-ray spectra for each EPIC camera from each observation is displayed in Fig.~\ref{fig:spec}. All of the spectra show a peak in emission around 0.6 to 1\,keV, except that for K2-95, which suffers from a lack of counts. This emission is chiefly produced by bound-bound transitions of highly-ionised Fe, and is typically stronger in younger stars. As a comparison, some of the older field stars (e.g. GJ\,436, HAT-P-11, and HD\,97658) we investigated in \citet[see fig. 3]{SurveyPaper}, show less pronounced emission at these energies, and have much softer spectra overall.

Overplotted on each spectra in Fig.~\ref{fig:spec} is our best fit model, as fitted using \textsc{xspec} 12.9.1p~\citep{Arnaud1996}. For the K2-100 observations, where we had a relatively large number of counts, we binned the spectra to a minimum of 25 counts prior to fitting. This choice was unsuitable for the other three objects, all of which had a more limited number of counts, and their spectra were therefore binned to a minimum of 10 counts. We used the C-statistic in fitting models to the spectra~\citep{Cash1979}. Throughout our analysis, we estimated uncertainties using a combination of \textsc{xspec}'s built-in MCMC sampler and error command. These values correspond to the 1-$\sigma$ (68 per cent) level, and are calculated from chains run for 100,000 steps following 5000 used for burn-in. Abundances were set to Solar values~\citep{Asplund2009}.

In all cases, APEC models were used to fit the spectrum~\citep{Smith2001}. These describe an isothermal, optically-thin plasma in collisional ionisation equilibrium. For K2-100, the spectra were of good enough quality to warrant a two-temperature fit. These temperatures were forced to be equal across the two observations for this star, but their normalisations were allowed to vary. The low count rates for both K2-104 and K2-95 meant that we had to limit the range of temperatures to be below 1.5\,keV, to prevent it reaching unreasonable values. 

Interstellar absorption was accounted for by including a multiplicative TBABS model term~\citep{Wilms2000}. To calculate the value of the hydrogen column density, \NH, we used a selective extinction, $E(B-V)$, for Praesepe of 0.027 \citep{Taylor2006}. Assuming the standard Milky Way reddening $R_V = A_V/E(B-V) = 3.1$ \citep{Schultz1975}, this gives a total extinction $A_V = 0.0837$. Finally, we use \NH\ = $2.21\times10^{21} A_V$ \citep{Guver2009} to give our final value of \NH = $1.85\times10^{20}$\,cm$^{-2}$, which we fixed in all of our fits.

Table~\ref{tab:fit+fluxes} summarises the findings of our spectral analysis of each observation. The best fit temperatures and emission measures are given, together with estimated fluxes, luminosities and planetary irradiation levels for each observation. The extrapolations to the EUV were performed using the empirical relations we derived in \citet{SurveyPaper}, based on the method of~\citet{Chadney2015}. 

The K2-100 data allows a comparison of the stellar X-ray output at two separate epochs. The flux is about 40-50\,\% higher in the 2013 data than the 2015 data. From both the spectra in Fig.~\ref{fig:spec} and the emission measures in Table~\ref{tab:fit+fluxes}, one can see that this change is being driven by the softest energies. The emission measures of the higher temperature component are in excellent agreement, but the 2015 emission measure for the lower temperature component is some five or six times smaller. 

All the results above are for the directly observable 0.2--2.4\,keV band. We also calculated fluxes for both \NH\ choices for each object in the 0.1--2.4\,keV band in \textsc{xspec}, in order to compare to previous studies of the rotation-X-ray output relationship. For these values we extrapolated the fitted model within \textsc{xspec}.

\section{\textit{K2} results}
\label{sec:K2}

\begin{figure*}
\centering
 \includegraphics[width=\textwidth]{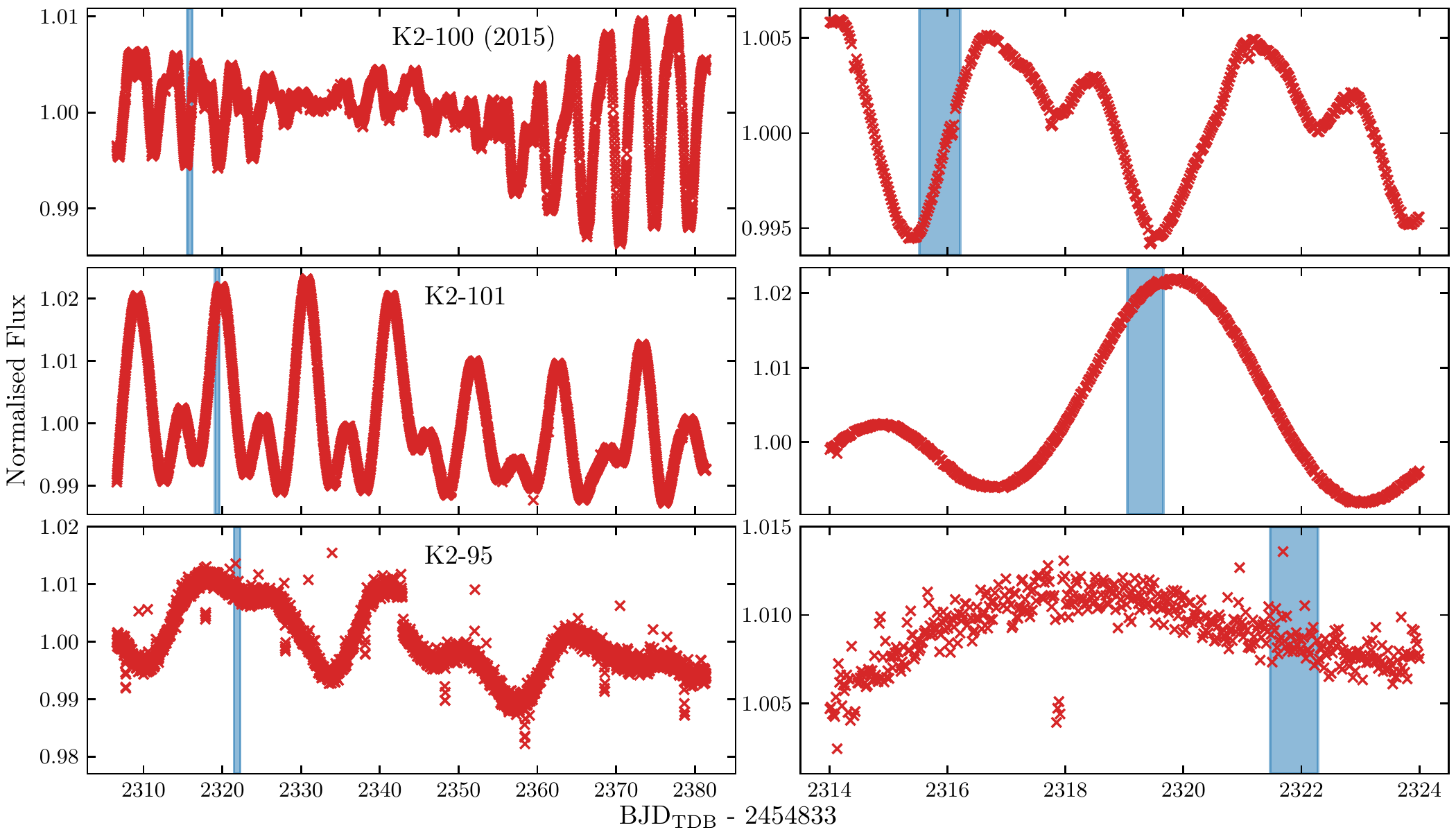}
 \caption[\textit{K2} light curves]{\textit{K2} light curves, detrended and corrected by the \textsc{everest} pipeline, for each of the three targets with simultaneous \textit{K2} and \textit{XMM-Newton} observations. Horizontally, the left-hand panels plot the full Campaign 5 light curves, while the right-hand panels zoom to the few days around where the simultaneous \textit{XMM-Newton} data were taken. Vertically, the top panels are for K2-100, the middle for K2-101, and the bottom for K2-95. The blue shaded region in all panels shows the epoch of \textit{XMM-Newton} observations for that system.}
 \label{fig:k2}
\end{figure*}

\begin{figure}
\centering
 \includegraphics[width=\columnwidth]{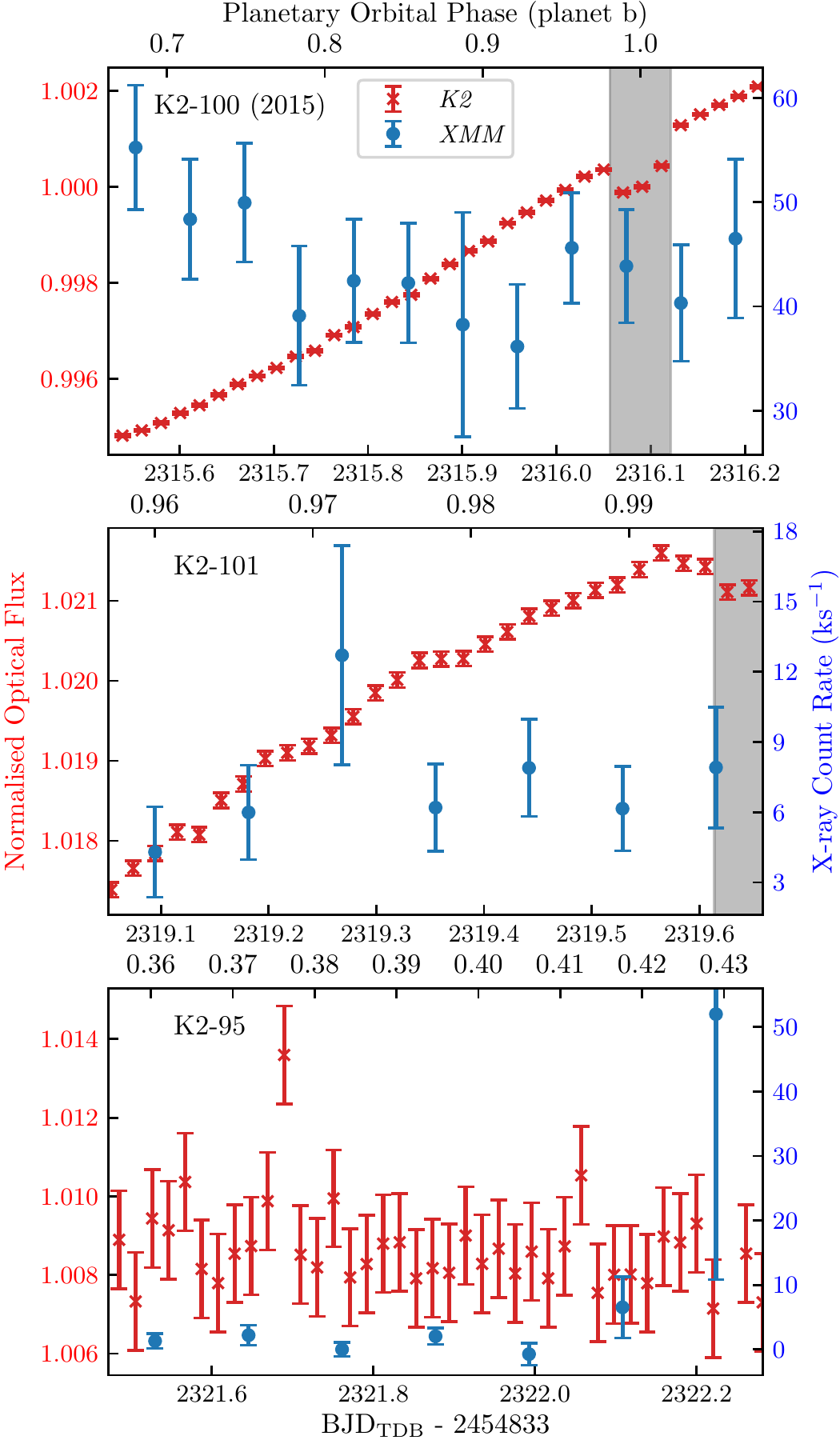}
 \caption[\textit{K2} and \textit{XMM-Newton} light curve comparison]{Comparison of the \textsc{everest} \textit{K2} and \textit{XMM-Newton} light curves. The top, middle, and bottom panels are for K2-100, K2-101, and K2-95, respectively. The \textit{K2} data are shown as red crosses, and the \textit{XMM-Newton} data are the blue circles. The grey shaded regions display the optical light transit epochs.}
 \label{fig:k2xmm}
\end{figure}

We examined the \textsc{everest}-corrected light curves for the three stars with simultaneous \textit{K2} and \textit{XMM-Newton} observations. This was performed with the aim of putting the X-ray observations into the context of the optical spot modulation, including searching for correlated variability between the optical and X-ray light curves. The data are plotted in Fig.~\ref{fig:k2}, which includes both the entire Campaign 5 light curve (left panels), and a zoomed in section (right panels) around the epoch of the \textit{XMM-Newton} observations, which is highlighted by the blue shaded region in each panel. All three stars show obvious spot modulation in their light curves, on the order of a few per cent.

From Fig.~\ref{fig:k2}, we can see that the K2-100 \textit{XMM-Newton} data cover a relatively large rise in the optical light curve, on the order of about 1 per cent. For K2-101, a much shallower rise in optical flux through the X-ray observation epoch is evident in the right-hand middle panel of Fig.~\ref{fig:k2}. However, the larger modulation amplitude for this star ($\sim$3 per cent versus $\sim$0.5--2 per cent for K2-100) means the fractional increase in the \textit{K2} flux during the \textit{XMM-Newton} observations is still about 0.4 per cent. The longer rotation period of K2-95 means that substantial changes in the optical flux due to spot modulation are on a timescale a little too long to be important through the length of an average \textit{XMM-Newton} observation. K2-95 also shows numerous outliers at elevated fluxes, perhaps suggestive of frequent flaring. However, such frequent flaring would not be so expected for a star with a 23.9\,d rotation period, and which does not show evidence of strong H$\alpha$ emission \citep{Douglas2014}.

In Fig.~\ref{fig:k2xmm}, we replot the \textit{XMM-Newton} light curves from Fig.~\ref{fig:xlc}, together with the simultaneous \textit{K2} data. This highlights an increase in the optical flux of K2-100 throughout the \textit{XMM-Newton} observation, simultaneous with a potential decrease in the X-ray flux followed by a plateau. Optically-dim starspots and X-ray bright coronal loops are both associated with active regions, and hence such behaviour of the X-ray and optical flux is reasonable and could be astrophysical. While K2-101 also shows a rise in optical flux through the time of the X-ray observations, albeit a smaller one fractionally, the factor of a few lower X-ray count rate likely means any effect on the X-ray output is too small to be seen within the uncertainties for this star.

\section{Discussion}
\label{sec:Discussion}

\subsection{Current XUV irradiation}
\label{ssec:Xoutput}

\begin{figure}
\centering
 \includegraphics[width=\columnwidth]{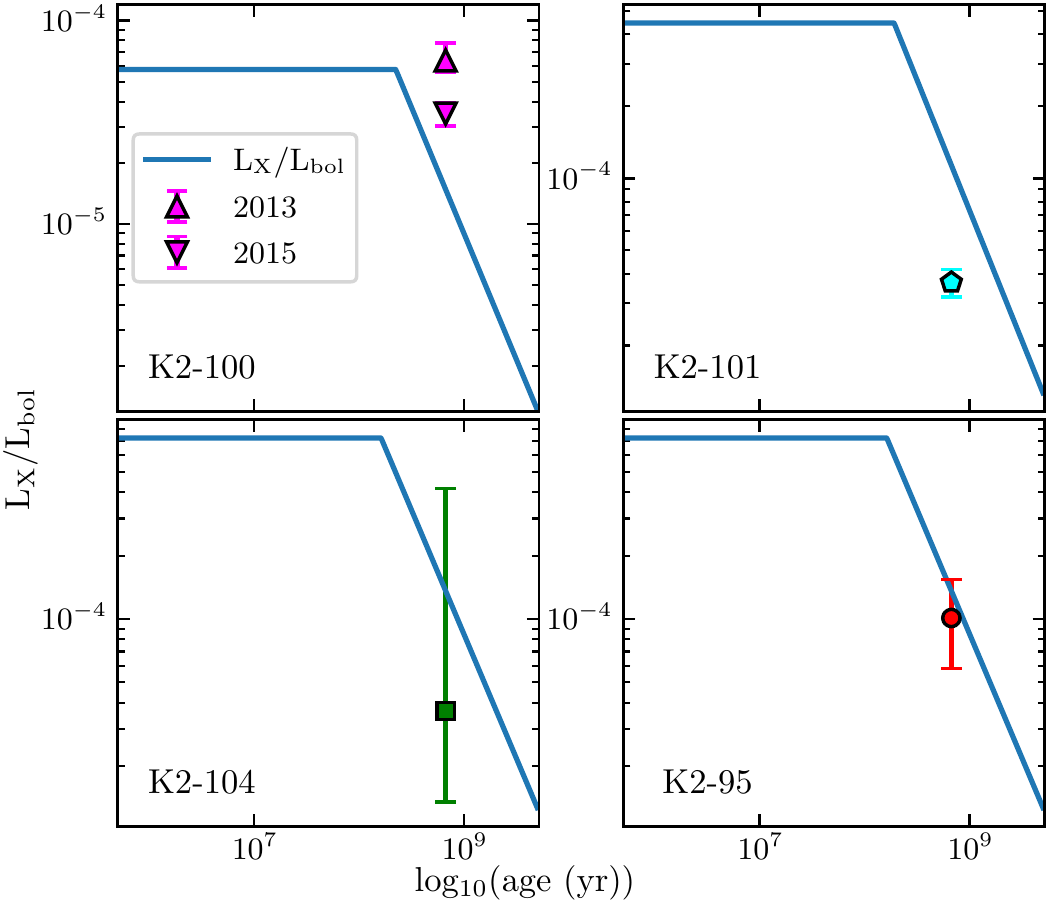}
 \caption{Comparison of the X-ray emission evolution of the stars in our sample, according the X-ray-age relations of J12. The measured $\log(L_{\rm X}/L_{\rm bol})$ in the 0.1--2.4\,keV band is also plotted for each star.}
 \label{fig:X_J12}
\end{figure}

\begin{figure}
 \includegraphics[width=\columnwidth]{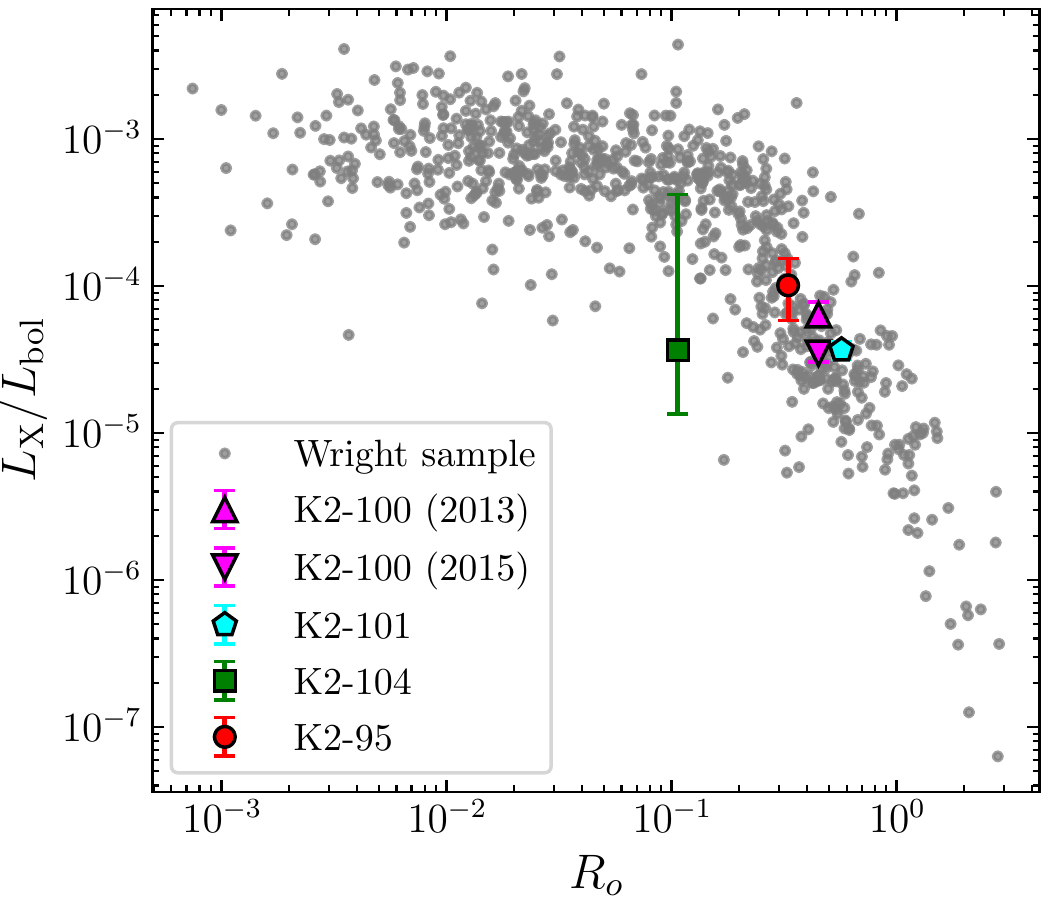}
 \caption{Replotting of fig.~1 of~\citet{Wright2016}, with points added for each of the five measurements presented in this work.}
 \label{fig:wrightSamp}
\end{figure}

Coronal X-ray emission reduces over a star's lifetime, as it spins down through magnetic braking. The relationship of X-ray output with both stellar age~\citep[e.g.][]{Guedel1997,Jackson2012,Nunez2016,Booth2017} and rotation period~\citep[e.g.][]{Pallavicini1981,Pizzolato2003,Wright2011,Douglas2014,Nunez2015,Wright2016,Wright2018} have been investigated extensively over the past few decades. For FGK stars, the emission is saturated for 100\,Myr or so, with \LxLb $\approx10^{-3}$, before falling off as a power law as the rotation period increases.

We give the $L_{\rm X}/L_{\rm bol}$ in the 0.1--2.4\,keV band for each observation in Table~\ref{tab:fit+fluxes}. To get $L_{\rm X}$ in the 0.1--2.4\,keV band, we extrapolated our fitted model within \textsc{xspec}. The values of $L_{\rm X}/L_{\rm bol}$ for this sample of host stars cluster around $10^{-5}$ to $10^{-4}$. These are all smaller, by an order of magnitude or more, than the observed saturation level at around $10^{-3}$. Therefore, the irradiation rates for the planets are now lower than they would have been over the first few hundred Myr of their lives, but still higher than around many middle-aged field stars.

This 0.1--2.4\,keV band is useful to us, as it allows a comparison to the X-ray-age relations of~\citet[hereafter, J12]{Jackson2012}. In Fig.~\ref{fig:X_J12}, the measured value of $L_{\rm X}/L_{\rm bol}$ is plotted together with the empirical relation for that star's colour bin in the J12 relations. K2-100 in particular has a considerably higher X-ray than may be expected from J12. This could be because the saturation level for the two bluest B-V colour bins in the J12 study is substantially lower than the rest. K2-100 falls into the second bluest, where the saturation level is $\log(L_{\rm X}/L_{\rm bol}) = -4.24$, far below the canonical value of about $-3$. Lower X-ray luminosity ratios for saturated stars bluer and more massive than the Sun have been found in other studies too \citep[e.g.][]{Pizzolato2003,Wright2011}, with the exact cause unknown. 

The 0.1--2.4\,keV band $L_{\rm X}/L_{\rm bol}$ also allow comparison with the X-ray-rotation works of \citet{Wright2011,Wright2016,Wright2018}. Our measurements are overplotted on their sample in Fig.~\ref{fig:wrightSamp}, wherein the Rossby number, $R_{\rm o}$, is given by $P_{\rm rot}/\tau$, where $\tau$ is the convective turnover time \citep{Noyes1984}. For K2-100, K2-101, and K2-95, we take values of $R_{\rm o}$ from N{\'u}{\~n}ez et al. (in prep). For K2-104, we take the known value of $P_{\rm rot}$ from M17 and combine it with an estimate of $\tau$ based on the $V-K_S$ colour using the relations of \citet{Wright2018} to estimate $R_{\rm o}$. All four stars have values that, within the uncertainties, are consistent with the wider sample plotted in Fig.~\ref{fig:wrightSamp}. K2-104's measured value is below the main sample and its $R_{\rm o}$, calculated via a different method, is lower. However, calculating $R_{\rm o}$ for the other three stars in the same way from \citet{Wright2018} (0.4202, 0.57, 0.2561 for K2-100, K2-101, and K2-95 respectively) gives values similar to those plotted from N{\'u}{\~n}ez et al. (in prep) (0.45, 0.57, 0.33), and still quite different compared to K2-104.

Overall, our observed values demonstrate the scatter around X-ray-age and X-ray-rotation relations and highlight the value of directly measuring X-ray fluxes, as opposed to solely relying on empirical relations where the scatter can be up to an order of magnitude.

\subsection{XUV irradiation evolution}
\label{ssec:irradEvo}

\begin{figure}
\centering
 \includegraphics[width=0.84\columnwidth]{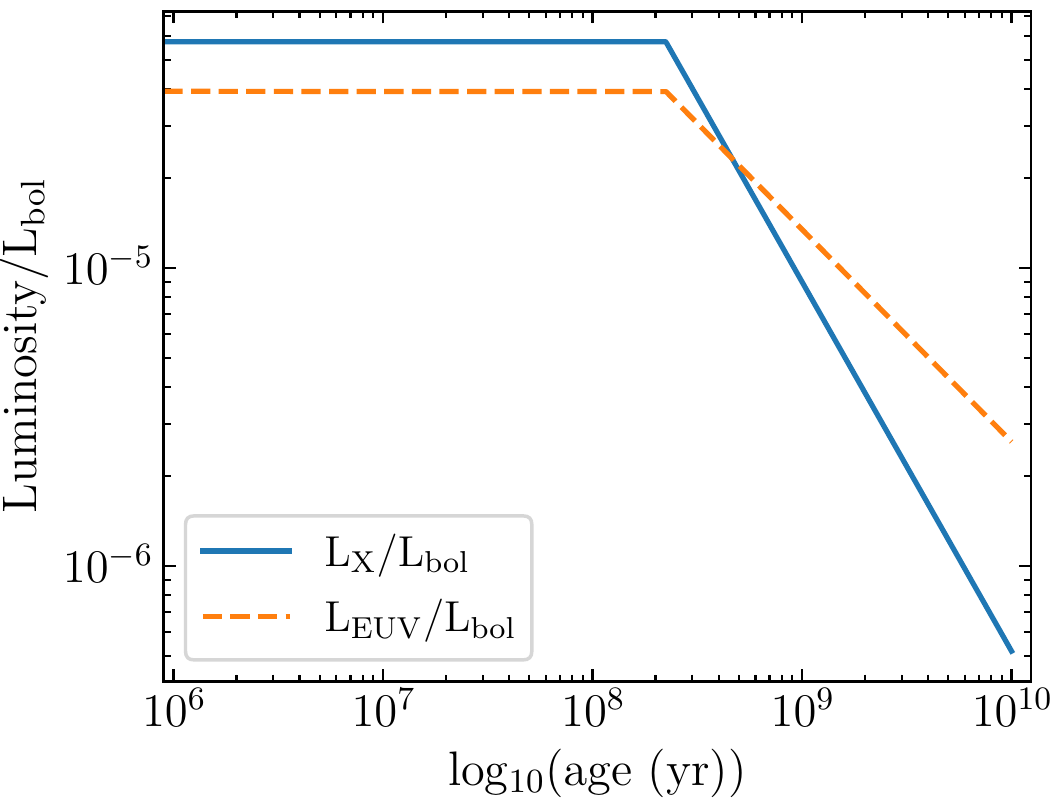}
 \caption[X-ray and EUV emission evolution]{Comparison of the X-ray (blue solid line) and EUV (orange dashed line) emission evolution expected for a star like K2-100. The X-ray evolution is that given by the X-ray-age relations of J12, and then extrapolated to EUV using the relations derived in \citet{SurveyPaper}.}
 \label{fig:XUV_J12}
\end{figure}

Given the high irradiation rate, it is often thought that the first few 100\,Myr is the most important epoch for the atmospheric evolution of close-in planets \citep[e.g.][]{LDE2007,Lopez2013,Johnstone2015,Owen2017,Livingston2018}. While this is the epoch of highest instantaneous irradiation, just as relevant is the gradient associated with the power law fall off of the XUV luminosity that follows. As long as this is more negative than -1, the saturated regime should dominate. In the J12 relations, the X-ray luminosity time evolution in the unsaturated regime is indeed steeper than this. However, in \citet{EUVevolution}, we combined the J12 relations with another set of relations for EUV reconstruction from X-rays \citep[derived using the method of \citealt{Chadney2015}]{SurveyPaper}. We found that the decline in EUV emission was far less steep, and in all cases the power law exponent was shallower than -1. The implication of this is that, unlike X-rays, the total integrated EUV per logarithmic decade of time actually increases. 
In Fig.~\ref{fig:XUV_J12}, we show how this implied shallower fall off of the EUV means it becomes the more dominant of the two bands over time, using a K2-100-like star as an example. Here, the EUV emission remains only an order of magnitude below the saturated regime at 5\,Gyr, while the X-rays has dropped off considerably more.


Integrating both the X-ray luminosity and extrapolated EUV time evolutions allowed us to calculate the total X-ray, EUV, and combined XUV irradiation across different stages of a star's lifetime. By way of example, in Table~\ref{tab:XUV_K100} we give the percentage of each energy band's radiated energy for a star like K2-100 for different epochs in its life, as compared to the total energy radiated between 1\,Myr and 10\,Gyr. Furthermore, in Table~\ref{tab:sampleXUVhist}, we compare the total irradiation in the different bands to date (age=670\,Myr) for the Praesepe planets in our sample, with that still to come, up to 10\,Gyr.

\begin{table}
  \caption{X-ray, EUV and combined XUV energy output by K2-100 at different epochs of its life, estimated using the relations of J12 together with the EUV extrapolations derived in \citet{SurveyPaper}. The percentages are as compared to the total energy output between 1\,Myr and 10\,Gyr.}
  \centering
  \label{tab:XUV_K100}
  \begin{threeparttable}
    \begin{tabular}{l c c c c c c}
    \hline\
    Epoch	& \multicolumn{2}{c}{X-ray} & \multicolumn{2}{c}{EUV} & \multicolumn{2}{c}{XUV} \\
    		& ($\dagger$) & (\%) & ($\dagger$) & (\%) & ($\dagger$) & (\%) \\
    \hline
    1--10\,Myr & 1.1 & 1.2 & 0.76 & 0.51 & 1.9 & 0.76 \\
    10--100\,Myr & 11 & 12 & 7.6 & 5.1 & 19 & 7.6 \\
    0.1--1\,Gyr & 50 & 52 & 46 & 31 & 96 & 39 \\
    1--10\,Gyr & 34 & 35 & 95 & 64 & 130 & 53 \\
    \hline
    1\,Myr--10\,Gyr & 96 & 100 & 150 & 100 & 250 & 100 \\ 
    \hline
\end{tabular}
\begin{tablenotes}
\item $\dagger$ $10^{44}$\,erg\ps
\end{tablenotes}
\end{threeparttable}
\end{table}

\begin{table}
  \caption{Past and future stellar emission, estimated using the relations of J12 together with the EUV extrapolations derived in \citet{SurveyPaper}. Given as \% of total output (1\,Myr to 10\,Gyr).}
  \centering
  \label{tab:sampleXUVhist}
    \begin{tabular}{l c c c c c c}
    \hline\
    Planet	& \multicolumn{2}{c}{X-ray} & \multicolumn{2}{c}{EUV} & \multicolumn{2}{c}{XUV} \\
    		& Past & Future & Past & Future & Past & Future \\
    \hline
    K2-100 & 59 & 41 & 31 & 69 & 42 & 58\\
    K2-101 & 53 & 47 & 28 & 72 & 40 & 60\\
    K2-104 & 60 & 40 & 31 & 69 & 43 & 57\\
    K2-95  & 60 & 40 & 31 & 69 & 42 & 58\\
    \hline
\end{tabular}
\end{table}

The values given in Table \ref{tab:XUV_K100} and \ref{tab:sampleXUVhist} have implications for not just the irradiation received by these stars' close-in planets over time, but also the atmospheric evolution of such planets. Some two-thirds of the total EUV emission expected for these stars up to 10\,Gyr is still to come, and even around two-fifths of the X-rays are in the future too. 
For mass loss under energy-limited assumptions (i.e. that the mass loss scales with the incident energy; see also Section~\ref{ssec:currentML}), these findings imply that significant mass loss could still occur far past the epoch where losses are thought to be large enough for appreciable atmospheric evolution. This is something we explore further in simulations of the future evolution of the planets in Section~\ref{ssec:Sims}. While some studies have explored the applicability and limitations of the energy-limited prescription \citep[e.g.][]{Owen2016,Erkaev2016,Krenn2021}, especially for smaller planets \citep[e.g.][]{Kubyshkina2018,Kubyshkina2018HBA}, its simplicity and ability to provide a plausible upper limit on the possible mass loss means it remains an attractive method of exploring mass loss. 

\subsection{Planetary mass loss}
\label{ssec:currentML}

\begin{table*}
\centering
  \caption{Measured radii and assumed masses for each planet, together with estimates of the current mass loss rate, using three methods: energy-limited escape ($\dot{M}_{\rm En}$), use of the interpolation tool across the \citet{Kubyshkina2018,Kubyshkina2021} grid of models ($\dot{M}_{\rm Kuby}$), and the \citet{Kubyshkina2018HBA} hydro-based approximation ($\dot{M}_{\rm HBA}$). Note for K2-100b, we use the mass measurement for all our mass loss calculations. The estimate from the mass-radius relation is provided for completeness and comparison only.}
  \label{tab:masslossEst}
  	\begin{threeparttable}
    \begin{tabular}{l c c c c c c c}
    \hline\
    Planet	& $R_{\rm p}$ & $M_{\rm p, meas}^\ast$ & $M_{\rm p, W16}^{\ast\ast}$ & $\log\dot{M}_{\rm En}$ & $\log\dot{M}_{\rm Kuby}$ & $\log\dot{M}_{\rm HBA}$ \\
    		& (R$_\oplus$)	& (M$_\oplus$) & (M$_\oplus$) & (g\,s$^{-1}$) & (g\,s$^{-1}$) \\
    \hline
    K2-100b (2013) & \multirow{2}{*}{$3.88\pm0.16$} & \multirow{2}{*}{$21.8\pm6.2$} & \multirow{2}{*}{$13.3^{+2.9}_{-2.8}$} & 11.3 & 11.4 & 12.7 \\
    K2-100b (2015) &  &  &  & 11.2 & 11.4 & 12.6 \\[0.1cm]
    K2-101b        & $2.0\pm0.1$ & - & $6.5\pm2.3$ & 9.0 & 9.2 & 9.0 \\[0.05cm]
    K2-104b        & $1.9^{+0.2}_{-0.1}$ & - & $6.1^{+2.5}_{-2.4}$ & 9.6 & 10.06 & 10.18 \\[0.05cm]
    K2-95b         & $3.7\pm0.2$ & - & $14.3^{+3.0}_{-2.9}$ & 9.5 & 9.5 & 8.8 \\[0.05cm]
    \hline
\end{tabular}
    \begin{tablenotes}
\item $^\ast$ Measured by \citet{Barragan2019}.
\item $^{\ast\ast}$ Estimated using the mass-radius relation of \citet{Wolfgang2016}.
    \end{tablenotes}
  \end{threeparttable}
\end{table*}

Using the XUV irradiation calculated in Section~\ref{ssec:XraySpec}, we were able to estimate the current mass loss rate for each planet. We initially adopted the energy-limited approximation \citep{Watson1981,Erkaev2007}, following numerous previous implementations in the literature~\citep[e.g.][]{LDE2007,SanzForcada2011,Salz2015,Louden2017,Wheatley2017,SurveyPaper}. Under this method, the mass loss rate, $\dot{M}_{\rm En}$, is given by
\begin{equation}
\dot{M}_{\rm En} = \frac{ \beta^2 \eta \pi F_{\rm XUV} R^3_{\rm p} }{ G K M_{\rm p} },
\label{eq:enML}
\end{equation}
where $\eta$ is the efficiency of the irradiation in driving mass loss, $\beta$ accounts for the height above the planetary radius that XUV photons are absorbed, and $K$ is a Roche lobe correction factor \citep{Erkaev2007}. Canonical values were assumed for $\eta$ and $\beta$ of 0.15 and 1.00. Of our sample, only K2-100 has a mass measurement \citep{Barragan2019}, and indeed this is the only radial velocity mass measurement to date for a planet in an open cluster. Such measurements are difficult to make, partially because of the higher activity level of the younger host stars. For the other three planets in our sample, we use the mass-radius relation and associated code\footnote{Their code: \url{https://github.com/dawolfgang/MRrelation}} of~\citet{Wolfgang2016} to estimate the mass, $M_{\rm p}$. These are given along with the resulting energy-limited mass loss rates in Table~\ref{tab:masslossEst}. For completeness and comparison, we include a mass estimate from \citet{Wolfgang2016} for K2-100b, but do not use this value in any of our mass loss estimates. We give two mass loss rates for K2-100b, corresponding to the 2013 and 2015 observations.

We also applied an interpolation tool that estimates mass loss according to the \citet{Kubyshkina2018} grid of models for low mass planets with hydrogen-dominated atmospheres \citep[see also][]{Kubyshkina2021}. These mass loss rates are given in Table \ref{tab:masslossEst}. We then also calculated mass loss rates using an analytic hydro-based approximation described in a follow up paper \citep{Kubyshkina2018HBA}. The values in this expression were attained by fitting it to the same original grid of models described in \citet{Kubyshkina2018}. We again give values calculated in this way for all of the planets in our sample in Table~\ref{tab:masslossEst}. Interestingly, given the methods are based on the same models, this method gives rates for K2-100b that are over an order of magnitude greater than the interpolator tool. This is despite the value from interpolation over the grid being a good match to that estimated from the energy-limited method. Such discrepancies between the interpolator and hydro-based approximation are most common for planets with small values of the Jeans escape parameter and orbital separation, for which Roche lobe position and stellar mass dependencies become relevant but are not considered in the approximation \citep{Kubyshkina2018HBA}. For K2-100b the Jeans escape parameter is not particularly small, but the planet does reside relatively close to its star. This, together with the somewhat high stellar mass within the range considered by the interpolator, could help to explain the discrepancy.

One should also note that if either of the smaller planets, K2-101b and K2-104b, do not retain a substantial hydrogen envelope, then the validity of both the energy-limited method and the \citet{Kubyshkina2018} models (which assume a hydrogen-dominated atmosphere) for those planets is uncertain.

K2-100b is almost certainly undergoing the greatest current level of mass loss, in absolute terms, of the planets in our sample. This is owing to the combination of its relatively large radius, together with its small separation from and large XUV output of its host star. If K2-104b still retains a substantial envelope, it is losing mass at only a slightly greater rate than K2-101b and K2-95, despite its much shorter orbital period and separation. For K2-101b, this is because the stellar XUV output 
is much greater than K2-104, while K2-95b has double the radius of K2-104b, and so the XUV absorbing region is about four times as large.


\subsection{Simulations of the planets' possible futures}
\label{ssec:Sims}

\begin{figure*}
\centering
 \includegraphics[width=0.88\textwidth]{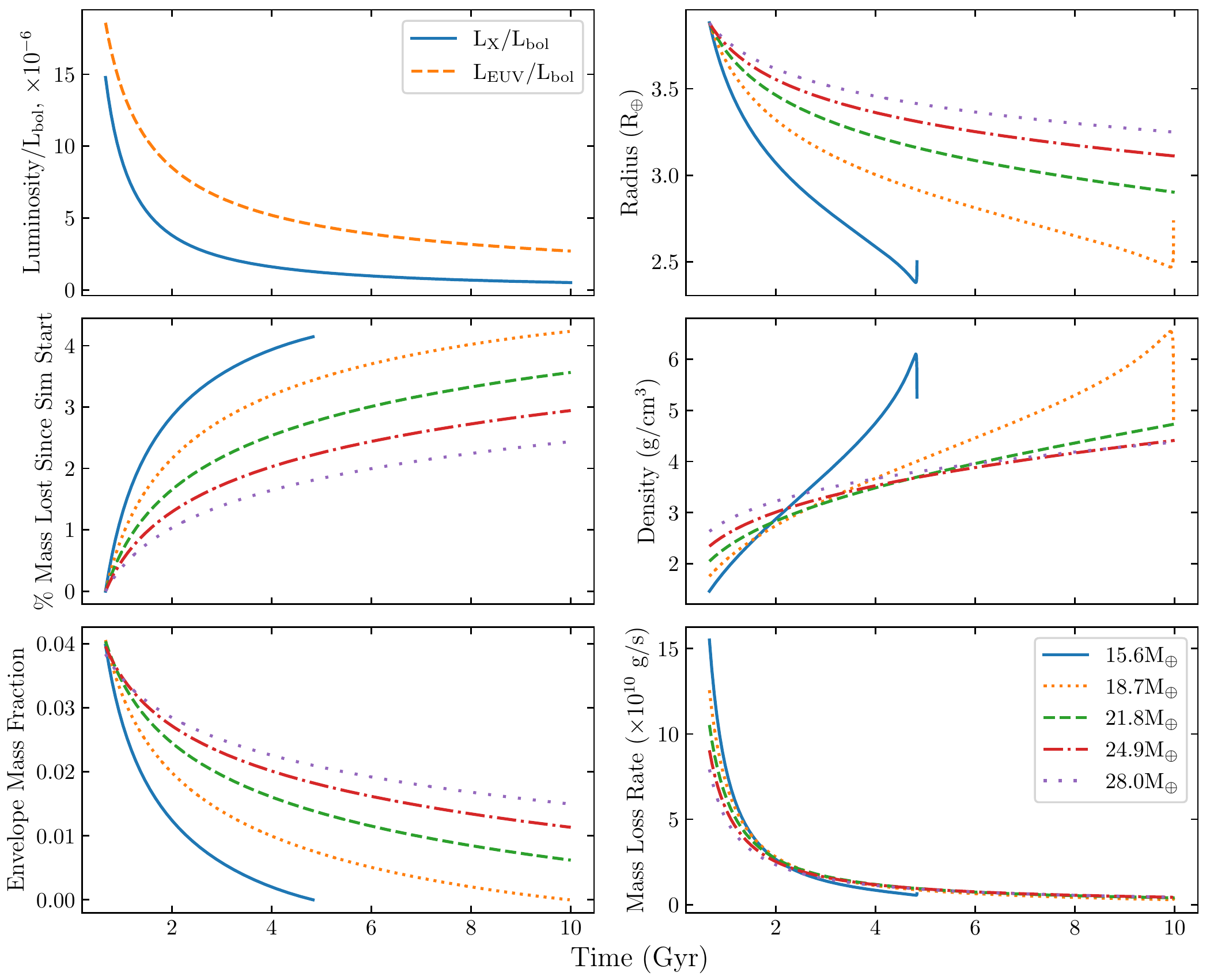}
 \caption{Results of our future simulations of mass loss from and its effects on the atmosphere of K2-100b. The top left panel shows the evolution of the X-ray and EUV luminosities in time. The other five panels show how six key values associated with the planet, and the escape of material from it evolve in time, starting at the current age of 750\,Myr. In each of these plots are a range of curves, each for a different choice of starting mass.}
 \label{fig:k2-100sims}
\end{figure*}

In \citet{EUVevolution}, we found that the time evolution of EUV emission drops off more slowly than for X-rays. The planets of Praesepe are still of relatively young age and so we explored the possible future evolution of the four planets in our sample by running some basic simulations. We investigated how the key properties of the planets - e.g. mass, radius, and density - may change in time.

To this end, we evolved forward each planet from its current age of 670\,Myr using 10\,kyr timesteps. For each planet, we started with a range of masses, listed in each planets relevant subsection (see Sections \ref{sssec:k2-100sim} through \ref{sssec:k2-95sims}). For each starting mass, we used equation~5 of \citet{Chen2016} to calculate the envelope mass fraction which recovered the measured planetary radius, as listed in Table~\ref{tab:SysParam}. We then modelled the XUV irradiation and resulting changes to the planet and its atmosphere. For the XUV time evolution, as we used in \citet{EUVevolution}, we took the X-ray-age relations of J12 for X-ray wavelengths, and applied the X-ray-EUV relations of \citet{SurveyPaper} to find the corresponding EUV time evolution. At each timestep, we then simply summed the two for the full XUV irradiation.

To evolve the planets at each timestep, we applied Equation \ref{eq:enML} for energy-limited mass loss to calculate the mass loss rate, again assuming canonical values of $\eta = 0.15$ and $\beta = 1$. The mass loss rate was then multiplied by the 10\,kyr timestep and the resulting amount of mass was removed from the envelope of the planet (we assumed that the core mass is constant across all ages). Using this, we calculated the planet mass and envelope mass fraction at the next timestep, and then used equation~5 of \citet{Chen2016} to find the corresponding new value of the radius. This entire process was repeated until either the age reached 10\,Gyr, or the envelope mass fraction reached zero. Throughout our calculations, we assumed that the values of $L_{\rm bol}$ and the semi-major axis did not change over the planets' lifetimes from their current measured values. We also ignore any other processes which may contribute to mass loss from the planets, such as core-powered mass loss. One should note that masses above 20\,M$_\oplus$, used for some of the K2-100b and K2-95b simulations, are outside of the limits for which the \citet{Chen2016} relationship was calculated. 
However, our calculations here are only intended to provide an illustration of the possible futures of these planets, and not a comprehensively modelled description.

\begin{figure*}
\centering
 \includegraphics[width=0.88\textwidth]{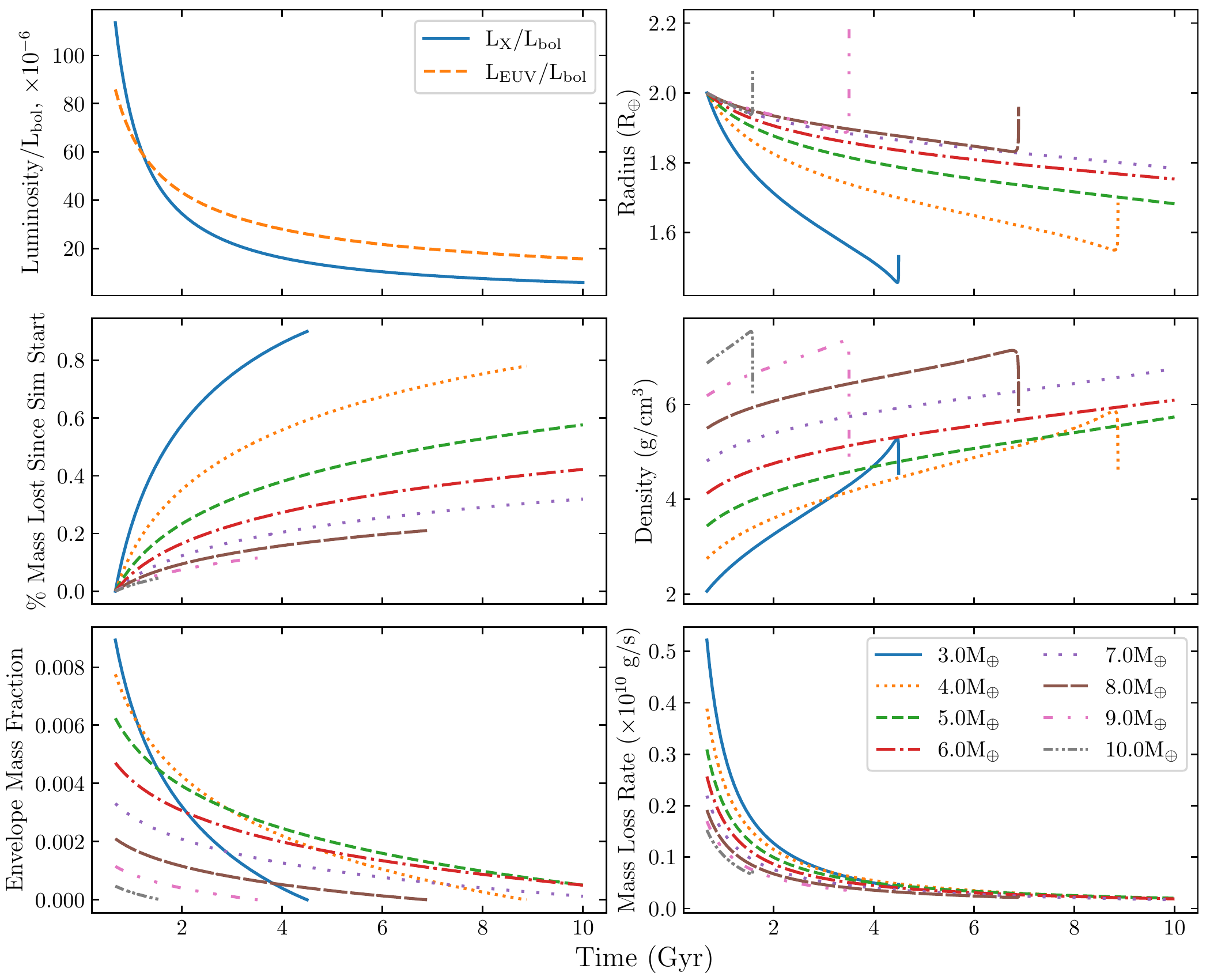}
 \caption{As Fig.~\ref{fig:k2-100sims}, but for K2-101b.}
 \label{fig:k2-101sims}
\end{figure*}

Our final results for each planet are displayed in Figs.~\ref{fig:k2-100sims} through \ref{fig:k2-95sims} for K2-100b, K2-101b, K2-104b, and K2-95b, respectively. For each planet, we have six subplots, detailing how the stellar XUV output, radius, total mass lost since the current age, density, envelope mass fraction, and mass loss rate are likely to evolve in the future. In the case of total evaporation, the simulation tracks tended to go vertical in radius (and thus density) at this end point. This is likely because the small amount of remaining gas expands to a much larger volume very quickly as the evaporation becomes a runaway process that the planet envelope can no longer prevent.

\subsubsection{K2-100b}
\label{sssec:k2-100sim}

K2-100b's position close to the Neptunian desert boundary (Fig.~\ref{fig:RpPorb})
makes it one of the more interesting planets to investigate with these simulations. K2-100b has a radial velocity mass measurement, providing starting point for the simulations. We chose to model values equal to this measurement, and 0.5 and 1 times the uncertainty either side (15.6, 18.7, 21.8, 24.9, and 28.0\,M$_\oplus$). In all cases, the calculated starting envelope mass fraction is about 4 per cent.

Our results for K2-100b are in Fig.~\ref{fig:k2-100sims}. Of the five mass choices we employed, the planetary atmosphere is stable across 10\,Gyr to complete evaporation in the three most massive cases. In the least massive case, in which it is fully stripped of its H/He envelope at an age of almost 6\,Gyr. The next least massive case (18.7\,M$_\oplus$) is fully stripped right as the age approaches 10\,Gyr. However, as the star is of G0V spectral type, it is expected that by around 10\,Gyr the host will have already reached at the end of its main sequence lifetime, and expansion of the star in the red giant phase could well occur first, engulfing K2-100b. 
For even the highest mass case, our results predict over half of the envelope that exists at the current age is lost by 10\,Gyr. 

In all cases the planet radius shrinks below 3.3\,$R_\oplus$, and in some cases considerably lower still, in line with evidence that that young planets in open clusters and stellar associations may be larger than their older counterparts \citep{Mann2017,Tofflemire2021}. Our results demonstrate that although K2-100b is currently close to or in the Neptunian desert, its position with respect to it could move down towards the boundary or somewhat below it by the time the system reaches an average field age of a few Gyr.

\subsubsection{K2-101b}

Unlike K2-100b, there exists no mass measurement of K2-101b. Instead, we used a range of masses that was centred approximately around the mass value estimated using \citet{Wolfgang2016} in Section~\ref{ssec:currentML} and displayed in Table~\ref{tab:masslossEst}. Specifically, we started the planet at integer multiples of Earth masses between 3 and 10\,M$_\oplus$ inclusive.

The results for K2-101b are displayed in Fig.~\ref{fig:k2-101sims}. The combination of the measured radius and our starting masses imply a range of starting envelope mass fraction between 0.05 and 1 per cent. Five of our eight test cases are stripped before 10\,Gyr. Three of these are the highest masses between 8 and 10\,M$_\oplus$, where it seems that the envelope mass fraction at the 670\,Myr start point was just too low for the envelope to survive. Conversely, the envelope mass fraction for the 3 and 4\,M$_\oplus$ cases were higher than the three where the envelope did survive. In this case, the core mass was likely too low to generate a deep enough gravitational well to hold on to the envelope across 10\,Gyr, and it is stripped by 4.5 and 9\,Gyr, respectively.

\begin{figure*}
\centering
 \includegraphics[width=0.88\textwidth]{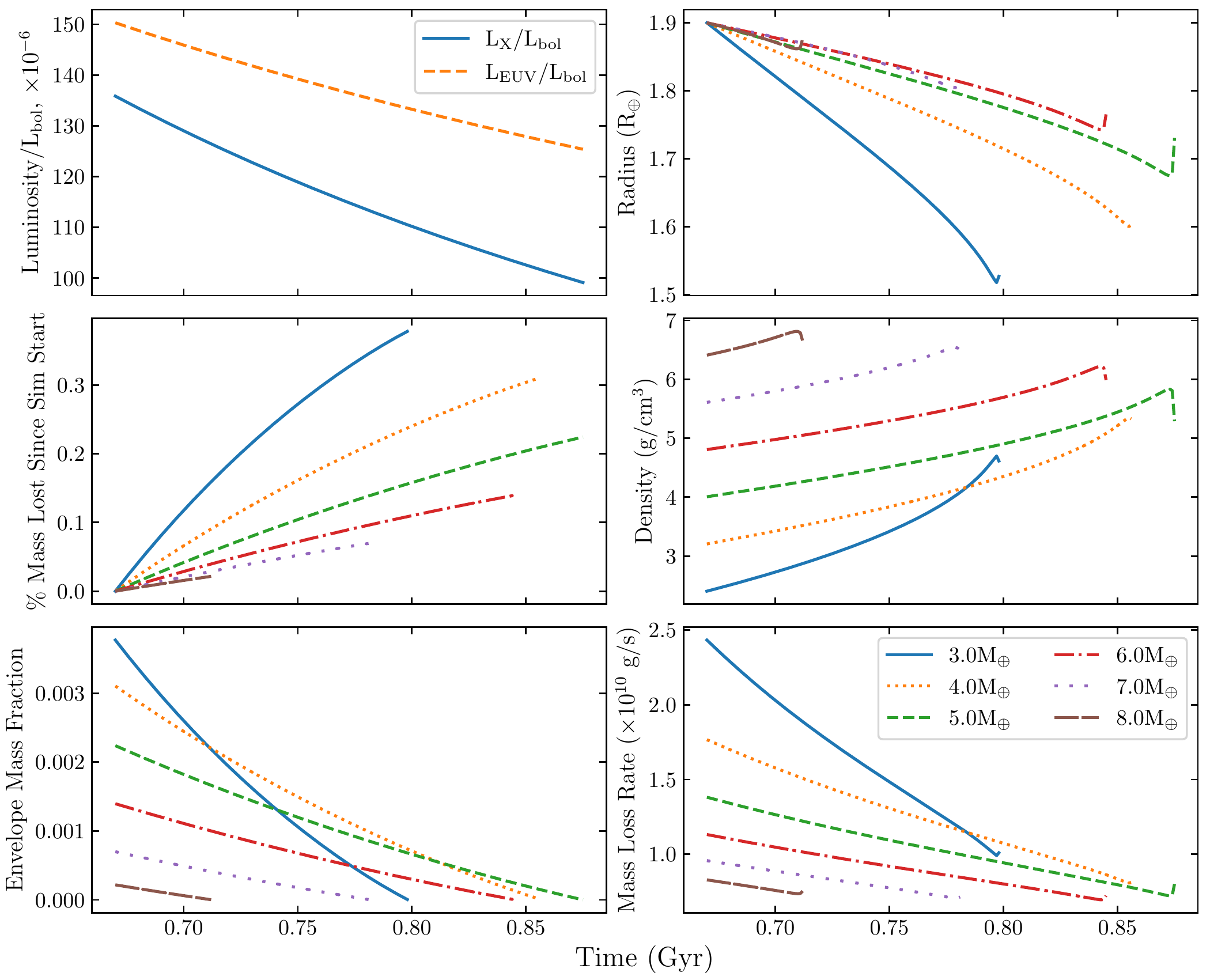}
 \caption{As Fig.~\ref{fig:k2-100sims}, but for K2-104b. Note the much smaller time scale as compared to the other plots.}
 \label{fig:k2-104sims}
\end{figure*}

The other four test cases with intermediate masses between 3 and 7\,M$_\oplus$ did successfully survive to 10\,Gyr, but the envelope mass fractions by this time suggest the planet would still be vulnerable to complete stripping later in its life (K2-101 is a K3V star, so its main-sequence lifetime will be considerably longer than the Sun or K2-100). The radii of these three cases by 10\,Gyr range from 1.68 to 1.78\,R$_\oplus$, smaller than the vast majority of known planets consistent with retaining a gaseous envelope, but still possibly above the radius-period valley. For a planet with an orbital period of 14.677\,d like K2-101b, the theoretical position of the valley is at 1.44\,R$_\oplus$ \citep{Owen2017}. Our conclusions for K2-101b are that if it does currently possess a small envelope of H/He, it could retain a portion of this for at least a substantial fraction of its lifetime, depending on its (core) mass.

\subsubsection{K2-104b}

There is no mass measurement of K2-104b. We therefore proceeded in similar fashion to K2-101, with starting masses between 3 and 8\,M$_\oplus$ inclusive. When we attempted to use masses greater than this, we found that no value of envelope mass fraction could recover the measured radius for the 9 or 10\,M$_\oplus$ cases.

Fig.~\ref{fig:k2-104sims} shows the results for K2-104b, where it should be noted that the time axis is different to the other three similar figures. This is because in all six test cases the envelope had been completely stripped, with the planet never retaining an envelope past the current age for very long. The most extreme case is the 8\,M$_\oplus$ test, which lasts only 42\,Myr into the simulations. The longest surviving case is 5\,M$_\oplus$, but it only lasts about 205\,Myr in, to an age of 875\,Myr. The existence of this turnover point in mass of the lifetime of the envelope in the simulations is likely similar to how the best surviving cases for K2-101b were those in the middle of the chosen mass range. At the high end of the mass range there is very little envelope for one to retain in the first place, so it is stripped off more quickly than the 5\,M$_\oplus$ case. At the low end of the mass range the planet is also stripped more quickly because the core mass is too low to hold on to the envelope for very long. 

An interesting, and outstanding question is whether the planet has an envelope to speak of at the current epoch. The theoretical radius-period valley of \citet{Owen2017} would suggest K2-104b is already below the valley, with any primordial light element envelope having already been stripped. The fact also that the highest envelope mass fraction of any of our test cases is just 0.4 per cent, with most much lower still, and that it does not last more than 205\,Myr into any of the simulations could be telling us that modelling it with an envelope at the current epoch is incorrect, particularly in the high core mass scenarios. A measurement of the mass, and thus bulk density, of the planet will prove difficult given the faintness of its M dwarf host star, but would help to shed light on whether any envelope exists at the current epoch. What is clear is that K2-104b will not retain a gaseous envelope by the time it is just a couple of Gyr old, with it either already having been stripped by its current age, or it is in the final stages of being so.

\subsubsection{K2-95b}
\label{sssec:k2-95sims}

\begin{figure*}
\centering
 \includegraphics[width=0.88\textwidth]{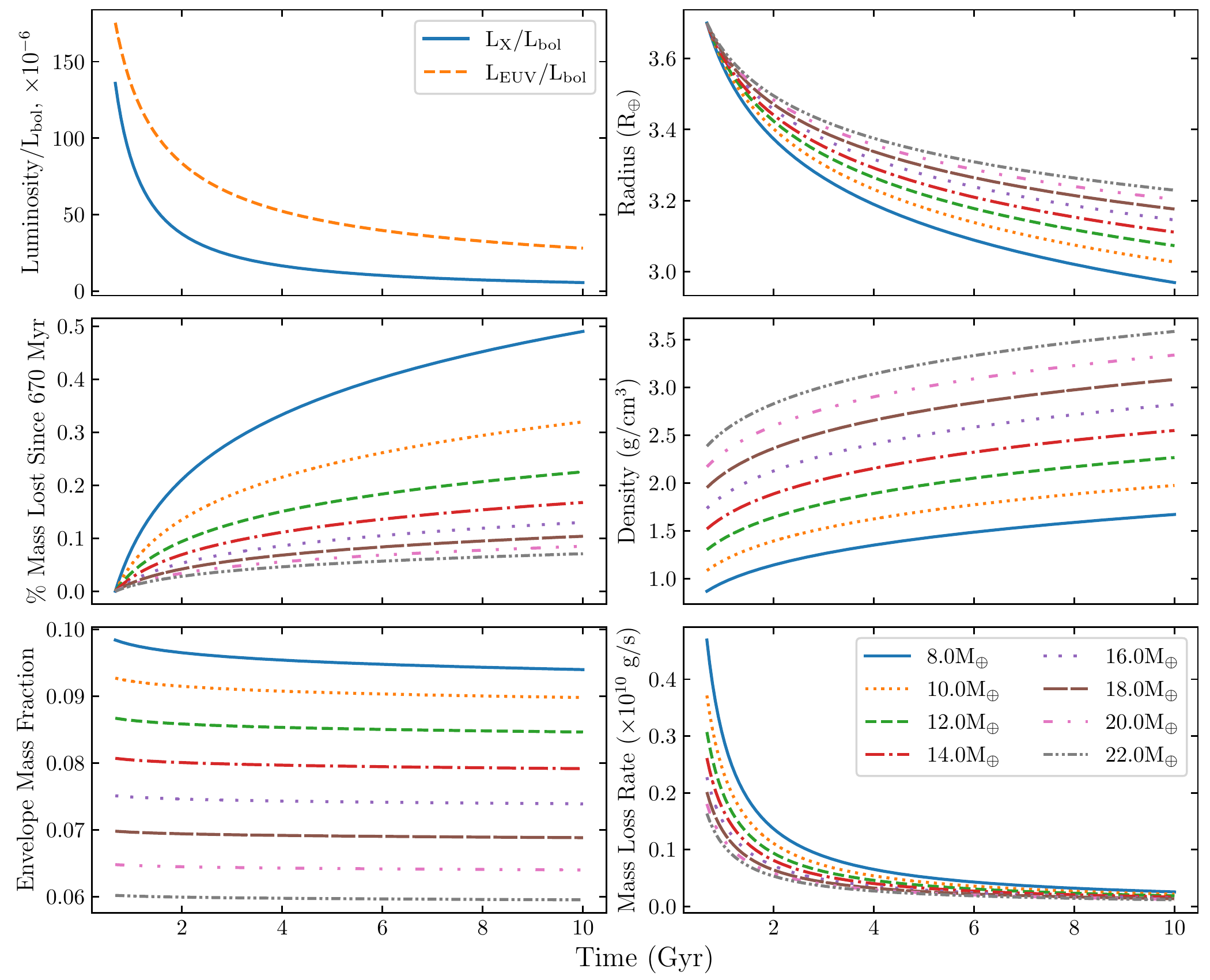}
 \caption{As Fig.~\ref{fig:k2-100sims}, but for K2-95b.}
 \label{fig:k2-95sims}
\end{figure*}

K2-95 also does not have a mass measurement, though the larger radius means that the \citet{Wolfgang2016} estimate of the mass is higher than for K2-101b or K2-104b. In this case, we started the planet with multiples of two Earth masses between 8 and 22\,M$_\oplus$ inclusive. 

Fig.~\ref{fig:k2-95sims} displays our results for K2-95b. This planet is clearly the most stable to evaporation in this small sample, with all of our test cases, spanning a wide range of low and intermediate bulk densities, easily surviving to an age of 10\,Gyr. The envelope mass fraction never drops below 6 per cent in any of the simulations, and in only one test, 8\,M$_\oplus$, does the planet drop to 3\,R$_\oplus$. That same test has by far the greatest mass loss of any of the simulations, with the planet losing 0.5 per cent of its mass at the current age by 10\,Gyr. However, this is less than a quarter of the lowest percentage mass loss in any of the K2-100b simulations. All of this likely comes about because of the combination of its larger size than K2-101b and K2-104b, and much its larger orbital period than K2-100b or K2-104b. In a more general context, the planet has a large enough radius that it is not close to the radius-period valley and so will not traverse it, and the planet has a long enough orbital period that it is not in the Neptunian desert under any definition at the current age.

\subsubsection{Alternative starting point informed by our \LxLb\ measurements}

In all of the simulations described above, we make use of the well-characterised age of the system as the starting point along the J12 high-energy time evolution tracks. As an alternative, we also used the measured values of \LxLb\ together with the J12 relations to estimate an ``X-ray-age" for each star (the age implied by the measured \LxLb\ for the 0.1--2.4\,keV band presented in Table \ref{tab:fit+fluxes}), which we then used as starting point on the J12 evolution tracks for another set of simulations. For K2-100 where we have two separate observations, we average its \LxLb\ across the two visits. The fact that these X-ray-ages are sometimes quite different from the known age of the system speaks to the scatter around relations such as those described in J12, which could be caused by short term or long term variability, and/or intrinsic star-to-star differences in the length of the saturation period and rate of the following decline.

When running these simulations, for K2-100b (X-ray-age for the system implied by measurement and J12: 252\,Myr), the 21.8\,M$_\oplus$ starting mass case additionally evaporates to a bare core, along with the two lighter cases that also evaporate in the 670\,Myr starting point simulations. The two most massive cases largely are unchanged, with their end points at 10\,Gyr slightly more compact and thus a bit denser. The choice of starting point seems to affect the outcomes for K2-101b (X-ray-age: 1.89\,Gyr). When starting at the known system age of 670\,Myr, five of the eight cases evaporate fully over the simulations to 10\,Gyr. With this X-ray-derived age, only the 9 and 10\,M$_\oplus$ cases evaporate fully, although it is questionable how much longer the 3\,M$_\oplus$ case would last past 10\,Gyr. For K2-104b (X-ray-age: 2.02\,Gyr) , all test cases still fully evaporate, just over a slightly longer time period because the XUV irradiation is lower at the higher age starting point. For K2-95 (X-ray-age: 861\,Myr), there is very little difference at all, since the X-ray-implied age is not so different from that known for the system, and this is the planet with the least change over the original simulations.

\section{Conclusions}
We have investigated four of the young planets discovered to date in the open cluster, Praesepe. X-ray analyses were performed for each of the four host stars using data taken by \textit{XMM-Newton}, allowing us to estimate the XUV irradiation experienced by planets at the current epoch. We found that although the XUV radiation is about an order of magnitude or so below the saturated level, the stars are still relatively active compared to older field stars, with considerable emission at energies of 0.6 to 1\,keV. The X-ray observations were paired with simultaneously taken \textit{K2} data, where possible, for further insight. For K2-100, the X-ray flux decreases and flattens off while the optical flux rises throughout, something that could result from active regions disappearing from view as the star spins.

We also estimated mass loss rates using a couple of different approaches, and found that K2-100b is undergoing the largest evaporation, in absolute terms, of the four at the current epoch. With our simulations of the future evolution, we find that K2-104b is likely to be stripped in the next few hundred Myr if it is not already. K2-100b will likely lose another few per cent of its current mass by 10\,Gyr, and could be susceptible to even being completely stripped of its envelope if its mass is at the lower end of the range allowed by RV measurements. K2-101b's future is sensitive to its current mass value, while K2-95b seems the most stable to evaporation processes.

Finally, discovery and characterisation of more planets with a range of ages up to a Gyr will help us better understand the mechanisms dominating the evolution of planetary atmospheres. X-ray observations of such systems are a key step in that characterisation process, allowing us to investigate the potentially escape-driving irradiation directly.

\section*{Acknowledgements}
We thank for the referee for their helpful comments and suggestions. PJW was supported by STFC consolidated grants ST/P000495/1 and ST/T000406/1. VAF acknowledges a quota studentship through grant code ST/S505365/1 funded by the STFC. NJM acknowledges support from STFC grant ST/S505444/1. MAA acknowledges support provided by NASA through grant 80NSSC21K0989. This work is based on observations obtained with {\it XMM-Newton}, an ESA science mission with instruments and contributions directly funded by ESA Member States and NASA. These observations are associated with OBSID 0721620101, 0761920901, 0761921001, and 0761921101. This paper includes data collected by the {\it K2} mission. Funding for the {\it K2} mission was provided by the NASA Science Mission directorate.

\section*{Data Availability}
The \textit{XMM-Newton} data used in this work are publically available from the \textit{XMM-Newton} Science Archive: \url{http://nxsa.esac.esa.int/nxsa-web/#search}. The \textit{K2} light curves produced by the \textsc{everest} pipeline are available through the Mikulski Archive for Space Telescopes (MAST): \url{https://archive.stsci.edu/k2/hlsp/everest/search.php}.



\bibliographystyle{mnras}
\bibliography{praesepe} 

\begin{thebibliography}{}
\makeatletter
\relax
\def\mn@urlcharsother{\let\do\@makeother \do\$\do\&\do\#\do\^\do\_\do\%\do\~}
\def\mn@doi{\begingroup\mn@urlcharsother \@ifnextchar [ {\mn@doi@}
  {\mn@doi@[]}}
\def\mn@doi@[#1]#2{\def\@tempa{#1}\ifx\@tempa\@empty \href
  {http://dx.doi.org/#2} {doi:#2}\else \href {http://dx.doi.org/#2} {#1}\fi
  \endgroup}
\def\mn@eprint#1#2{\mn@eprint@#1:#2::\@nil}
\def\mn@eprint@arXiv#1{\href {http://arxiv.org/abs/#1} {{\tt arXiv:#1}}}
\def\mn@eprint@dblp#1{\href {http://dblp.uni-trier.de/rec/bibtex/#1.xml}
  {dblp:#1}}
\def\mn@eprint@#1:#2:#3:#4\@nil{\def\@tempa {#1}\def\@tempb {#2}\def\@tempc
  {#3}\ifx \@tempc \@empty \let \@tempc \@tempb \let \@tempb \@tempa \fi \ifx
  \@tempb \@empty \def\@tempb {arXiv}\fi \@ifundefined
  {mn@eprint@\@tempb}{\@tempb:\@tempc}{\expandafter \expandafter \csname
  mn@eprint@\@tempb\endcsname \expandafter{\@tempc}}}

\bibitem[\protect\citeauthoryear{{Arnaud}}{{Arnaud}}{1996}]{Arnaud1996}
{Arnaud} K.~A.,  1996, in {Jacoby} G.~H.,  {Barnes} J.,  eds,  Astronomical
  Society of the Pacific Conference Series Vol. 101, Astronomical Data Analysis
  Software and Systems V. p.~17

\bibitem[\protect\citeauthoryear{{Asplund}, {Grevesse}, {Sauval}  \&
  {Scott}}{{Asplund} et~al.}{2009}]{Asplund2009}
{Asplund} M.,  {Grevesse} N.,  {Sauval} A.~J.,   {Scott} P.,  2009, \mn@doi
  [\araa] {10.1146/annurev.astro.46.060407.145222}, \href
  {http://adsabs.harvard.edu/abs/2009ARA%26A..47..481A} {47, 481}

\bibitem[\protect\citeauthoryear{{Barrag{\'a}n} et~al.,}{{Barrag{\'a}n}
  et~al.}{2019}]{Barragan2019}
{Barrag{\'a}n} O.,  et~al., 2019, \mn@doi [\mnras] {10.1093/mnras/stz2569},
  \href {https://ui.adsabs.harvard.edu/abs/2019MNRAS.490..698B} {490, 698}

\bibitem[\protect\citeauthoryear{{Barros}, {Demangeon}  \& {Deleuil}}{{Barros}
  et~al.}{2016}]{Barros2016}
{Barros} S.~C.~C.,  {Demangeon} O.,   {Deleuil} M.,  2016, \mn@doi [\aap]
  {10.1051/0004-6361/201628902}, \href
  {http://adsabs.harvard.edu/abs/2016A%26A...594A.100B} {594, A100}

\bibitem[\protect\citeauthoryear{{Batalha} et~al.,}{{Batalha}
  et~al.}{2010}]{Batalha2010}
{Batalha} N.~M.,  et~al., 2010, \mn@doi [\apjl] {10.1088/2041-8205/713/2/L109},
  \href {http://adsabs.harvard.edu/abs/2010ApJ...713L.109B} {713, L109}

\bibitem[\protect\citeauthoryear{{Beaug{\'e}} \& {Nesvorn{\'y}}}{{Beaug{\'e}}
  \& {Nesvorn{\'y}}}{2013}]{Beauge2013}
{Beaug{\'e}} C.,  {Nesvorn{\'y}} D.,  2013, \mn@doi [\apj]
  {10.1088/0004-637X/763/1/12}, \href
  {http://adsabs.harvard.edu/abs/2013ApJ...763...12B} {763, 12}

\bibitem[\protect\citeauthoryear{{Booth}, {Poppenhaeger}, {Watson}, {Silva
  Aguirre}  \& {Wolk}}{{Booth} et~al.}{2017}]{Booth2017}
{Booth} R.~S.,  {Poppenhaeger} K.,  {Watson} C.~A.,  {Silva Aguirre} V.,
  {Wolk} S.~J.,  2017, \mn@doi [\mnras] {10.1093/mnras/stx1630}, \href
  {http://adsabs.harvard.edu/abs/2017MNRAS.471.1012B} {471, 1012}

\bibitem[\protect\citeauthoryear{{Cash}}{{Cash}}{1979}]{Cash1979}
{Cash} W.,  1979, \mn@doi [\apj] {10.1086/156922}, \href
  {http://adsabs.harvard.edu/abs/1979ApJ...228..939C} {228, 939}

\bibitem[\protect\citeauthoryear{{Chadney}, {Galand}, {Unruh}, {Koskinen}  \&
  {Sanz-Forcada}}{{Chadney} et~al.}{2015}]{Chadney2015}
{Chadney} J.~M.,  {Galand} M.,  {Unruh} Y.~C.,  {Koskinen} T.~T.,
  {Sanz-Forcada} J.,  2015, \mn@doi [\icarus] {10.1016/j.icarus.2014.12.012},
  \href {http://adsabs.harvard.edu/abs/2015Icar..250..357C} {250, 357}

\bibitem[\protect\citeauthoryear{{Chen} \& {Rogers}}{{Chen} \&
  {Rogers}}{2016}]{Chen2016}
{Chen} H.,  {Rogers} L.~A.,  2016, \mn@doi [\apj]
  {10.3847/0004-637X/831/2/180}, \href
  {http://adsabs.harvard.edu/abs/2016ApJ...831..180C} {831, 180}

\bibitem[\protect\citeauthoryear{{Douglas} et~al.,}{{Douglas}
  et~al.}{2014}]{Douglas2014}
{Douglas} S.~T.,  et~al., 2014, \mn@doi [\apj] {10.1088/0004-637X/795/2/161},
  \href {https://ui.adsabs.harvard.edu/abs/2014ApJ...795..161D} {795, 161}

\bibitem[\protect\citeauthoryear{{Douglas}, {Curtis}, {Ag{\"u}eros}, {Cargile},
  {Brewer}, {Meibom}  \& {Jansen}}{{Douglas} et~al.}{2019}]{Douglas2019}
{Douglas} S.~T.,  {Curtis} J.~L.,  {Ag{\"u}eros} M.~A.,  {Cargile} P.~A.,
  {Brewer} J.~M.,  {Meibom} S.,   {Jansen} T.,  2019, \mn@doi [\apj]
  {10.3847/1538-4357/ab2468}, \href
  {https://ui.adsabs.harvard.edu/abs/2019ApJ...879..100D} {879, 100}

\bibitem[\protect\citeauthoryear{{Erkaev}, {Kulikov}, {Lammer}, {Selsis},
  {Langmayr}, {Jaritz}  \& {Biernat}}{{Erkaev} et~al.}{2007}]{Erkaev2007}
{Erkaev} N.~V.,  {Kulikov} Y.~N.,  {Lammer} H.,  {Selsis} F.,  {Langmayr} D.,
  {Jaritz} G.~F.,   {Biernat} H.~K.,  2007, \mn@doi [\aap]
  {10.1051/0004-6361:20066929}, \href
  {https://ui.adsabs.harvard.edu/abs/2007A&A...472..329E} {472, 329}

\bibitem[\protect\citeauthoryear{{Erkaev}, {Lammer}, {Odert}, {Kislyakova},
  {Johnstone}, {G{\"u}del}  \& {Khodachenko}}{{Erkaev}
  et~al.}{2016}]{Erkaev2016}
{Erkaev} N.~V.,  {Lammer} H.,  {Odert} P.,  {Kislyakova} K.~G.,  {Johnstone}
  C.~P.,  {G{\"u}del} M.,   {Khodachenko} M.~L.,  2016, \mn@doi [\mnras]
  {10.1093/mnras/stw935}, \href
  {https://ui.adsabs.harvard.edu/abs/2016MNRAS.460.1300E} {460, 1300}

\bibitem[\protect\citeauthoryear{{Feigelson} et~al.,}{{Feigelson}
  et~al.}{2004}]{Feigelson2004}
{Feigelson} E.~D.,  et~al., 2004, \mn@doi [\apj] {10.1086/422248}, \href
  {http://adsabs.harvard.edu/abs/2004ApJ...611.1107F} {611, 1107}

\bibitem[\protect\citeauthoryear{{Fulton} \& {Petigura}}{{Fulton} \&
  {Petigura}}{2018}]{Fulton2018}
{Fulton} B.~J.,  {Petigura} E.~A.,  2018, \mn@doi [\aj]
  {10.3847/1538-3881/aae828}, \href
  {https://ui.adsabs.harvard.edu/abs/2018AJ....156..264F} {156, 264}

\bibitem[\protect\citeauthoryear{{Fulton} et~al.,}{{Fulton}
  et~al.}{2017}]{Fulton2017}
{Fulton} B.~J.,  et~al., 2017, \mn@doi [\aj] {10.3847/1538-3881/aa80eb}, \href
  {http://adsabs.harvard.edu/abs/2017AJ....154..109F} {154, 109}

\bibitem[\protect\citeauthoryear{{Gaia Collaboration} et~al.,}{{Gaia
  Collaboration} et~al.}{2021}]{GaiaEDR3}
{Gaia Collaboration} et~al., 2021, \mn@doi [\aap]
  {10.1051/0004-6361/202039657}, \href
  {https://ui.adsabs.harvard.edu/abs/2021A&A...649A...1G} {649, A1}

\bibitem[\protect\citeauthoryear{{Ginzburg}, {Schlichting}  \&
  {Sari}}{{Ginzburg} et~al.}{2016}]{Ginzburg2016}
{Ginzburg} S.,  {Schlichting} H.~E.,   {Sari} R.,  2016, \mn@doi [\apj]
  {10.3847/0004-637X/825/1/29}, \href
  {https://ui.adsabs.harvard.edu/abs/2016ApJ...825...29G} {825, 29}

\bibitem[\protect\citeauthoryear{{G{\"u}del}, {Guinan}  \&
  {Skinner}}{{G{\"u}del} et~al.}{1997}]{Guedel1997}
{G{\"u}del} M.,  {Guinan} E.~F.,   {Skinner} S.~L.,  1997, \mn@doi [\apj]
  {10.1086/304264}, \href {http://adsabs.harvard.edu/abs/1997ApJ...483..947G}
  {483, 947}

\bibitem[\protect\citeauthoryear{{Gupta} \& {Schlichting}}{{Gupta} \&
  {Schlichting}}{2019}]{Gupta2019}
{Gupta} A.,  {Schlichting} H.~E.,  2019, \mn@doi [\mnras]
  {10.1093/mnras/stz1230}, \href
  {https://ui.adsabs.harvard.edu/abs/2019MNRAS.487...24G} {487, 24}

\bibitem[\protect\citeauthoryear{{Gupta} \& {Schlichting}}{{Gupta} \&
  {Schlichting}}{2020}]{Gupta2020}
{Gupta} A.,  {Schlichting} H.~E.,  2020, \mn@doi [\mnras]
  {10.1093/mnras/staa315}, \href
  {https://ui.adsabs.harvard.edu/abs/2020MNRAS.493..792G} {493, 792}

\bibitem[\protect\citeauthoryear{{G{\"u}ver} \& {{\"O}zel}}{{G{\"u}ver} \&
  {{\"O}zel}}{2009}]{Guver2009}
{G{\"u}ver} T.,  {{\"O}zel} F.,  2009, \mn@doi [\mnras]
  {10.1111/j.1365-2966.2009.15598.x}, \href
  {https://ui.adsabs.harvard.edu/abs/2009MNRAS.400.2050G} {400, 2050}

\bibitem[\protect\citeauthoryear{{Helled}, {Lozovsky}  \& {Zucker}}{{Helled}
  et~al.}{2016}]{Helled2016}
{Helled} R.,  {Lozovsky} M.,   {Zucker} S.,  2016, \mn@doi [\mnras]
  {10.1093/mnrasl/slv158}, \href
  {http://adsabs.harvard.edu/abs/2016MNRAS.455L..96H} {455, L96}

\bibitem[\protect\citeauthoryear{{Henden}, {Levine}, {Terrell}, {Smith}  \&
  {Welch}}{{Henden} et~al.}{2012}]{Henden2012}
{Henden} A.~A.,  {Levine} S.~E.,  {Terrell} D.,  {Smith} T.~C.,   {Welch} D.,
  2012, Journal of the American Association of Variable Star Observers
  (JAAVSO), \href {http://adsabs.harvard.edu/abs/2012JAVSO..40..430H} {40, 430}

\bibitem[\protect\citeauthoryear{{Howell} et~al.,}{{Howell}
  et~al.}{2014}]{Howell2014}
{Howell} S.~B.,  et~al., 2014, \mn@doi [\pasp] {10.1086/676406}, \href
  {https://ui.adsabs.harvard.edu/abs/2014PASP..126..398H} {126, 398}

\bibitem[\protect\citeauthoryear{{Jackson}, {Davis}  \& {Wheatley}}{{Jackson}
  et~al.}{2012}]{Jackson2012}
{Jackson} A.~P.,  {Davis} T.~A.,   {Wheatley} P.~J.,  2012, \mn@doi [\mnras]
  {10.1111/j.1365-2966.2012.20657.x}, \href
  {http://adsabs.harvard.edu/abs/2012MNRAS.422.2024J} {422, 2024}

\bibitem[\protect\citeauthoryear{{Jansen} et~al.,}{{Jansen}
  et~al.}{2001}]{Jansen2001}
{Jansen} F.,  et~al., 2001, \mn@doi [\aap] {10.1051/0004-6361:20000036}, \href
  {https://ui.adsabs.harvard.edu/abs/2001A&A...365L...1J} {365, L1}

\bibitem[\protect\citeauthoryear{{Jin} \& {Mordasini}}{{Jin} \&
  {Mordasini}}{2018}]{Jin2018}
{Jin} S.,  {Mordasini} C.,  2018, \mn@doi [\apj] {10.3847/1538-4357/aa9f1e},
  \href {https://ui.adsabs.harvard.edu/abs/2018ApJ...853..163J} {853, 163}

\bibitem[\protect\citeauthoryear{{Jin}, {Mordasini}, {Parmentier}, {van
  Boekel}, {Henning}  \& {Ji}}{{Jin} et~al.}{2014}]{Jin2014}
{Jin} S.,  {Mordasini} C.,  {Parmentier} V.,  {van Boekel} R.,  {Henning} T.,
  {Ji} J.,  2014, \mn@doi [\apj] {10.1088/0004-637X/795/1/65}, \href
  {http://adsabs.harvard.edu/abs/2014ApJ...795...65J} {795, 65}

\bibitem[\protect\citeauthoryear{{Johnstone} et~al.,}{{Johnstone}
  et~al.}{2015}]{Johnstone2015}
{Johnstone} C.~P.,  et~al., 2015, \mn@doi [\apjl]
  {10.1088/2041-8205/815/1/L12}, \href
  {https://ui.adsabs.harvard.edu/abs/2015ApJ...815L..12J} {815, L12}

\bibitem[\protect\citeauthoryear{{Jordi}, {Grebel}  \& {Ammon}}{{Jordi}
  et~al.}{2006}]{Jordi2006}
{Jordi} K.,  {Grebel} E.~K.,   {Ammon} K.,  2006, \mn@doi [\aap]
  {10.1051/0004-6361:20066082}, \href
  {http://adsabs.harvard.edu/abs/2006A%26A...460..339J} {460, 339}

\bibitem[\protect\citeauthoryear{{King} \& {Wheatley}}{{King} \&
  {Wheatley}}{2021}]{EUVevolution}
{King} G.~W.,  {Wheatley} P.~J.,  2021, \mn@doi [\mnras]
  {10.1093/mnrasl/slaa186}, \href
  {https://ui.adsabs.harvard.edu/abs/2021MNRAS.501L..28K} {501, L28}

\bibitem[\protect\citeauthoryear{{King} et~al.,}{{King}
  et~al.}{2018}]{SurveyPaper}
{King} G.~W.,  et~al., 2018, \mn@doi [\mnras] {10.1093/mnras/sty1110}, \href
  {http://adsabs.harvard.edu/abs/2018MNRAS.tmp.1058K} {478, 1193}

\bibitem[\protect\citeauthoryear{{Krenn}, {Fossati}, {Kubyshkina}  \&
  {Lammer}}{{Krenn} et~al.}{2021}]{Krenn2021}
{Krenn} A.~F.,  {Fossati} L.,  {Kubyshkina} D.,   {Lammer} H.,  2021, \mn@doi
  [\aap] {10.1051/0004-6361/202140437}, \href
  {https://ui.adsabs.harvard.edu/abs/2021A&A...650A..94K} {650, A94}

\bibitem[\protect\citeauthoryear{{Kubyshkina} \& {Fossati}}{{Kubyshkina} \&
  {Fossati}}{2021}]{Kubyshkina2021}
{Kubyshkina} D.~I.,  {Fossati} L.,  2021, \mn@doi [Research Notes of the
  American Astronomical Society] {10.3847/2515-5172/abf498}, \href
  {https://ui.adsabs.harvard.edu/abs/2021RNAAS...5...74K} {5, 74}

\bibitem[\protect\citeauthoryear{{Kubyshkina} et~al.,}{{Kubyshkina}
  et~al.}{2018a}]{Kubyshkina2018}
{Kubyshkina} D.,  et~al., 2018a, \mn@doi [\aap] {10.1051/0004-6361/201833737},
  \href {https://ui.adsabs.harvard.edu/abs/2018A&A...619A.151K} {619, A151}

\bibitem[\protect\citeauthoryear{{Kubyshkina} et~al.,}{{Kubyshkina}
  et~al.}{2018b}]{Kubyshkina2018HBA}
{Kubyshkina} D.,  et~al., 2018b, \mn@doi [\apjl] {10.3847/2041-8213/aae586},
  \href {https://ui.adsabs.harvard.edu/abs/2018ApJ...866L..18K} {866, L18}

\bibitem[\protect\citeauthoryear{{Kurokawa} \& {Nakamoto}}{{Kurokawa} \&
  {Nakamoto}}{2014}]{Kurokawa2014}
{Kurokawa} H.,  {Nakamoto} T.,  2014, \mn@doi [\apj]
  {10.1088/0004-637X/783/1/54}, \href
  {http://adsabs.harvard.edu/abs/2014ApJ...783...54K} {783, 54}

\bibitem[\protect\citeauthoryear{{Lecavelier Des Etangs}}{{Lecavelier Des
  Etangs}}{2007}]{LDE2007}
{Lecavelier Des Etangs} A.,  2007, \mn@doi [\aap] {10.1051/0004-6361:20065014},
  \href {http://adsabs.harvard.edu/abs/2007A%26A...461.1185L} {461, 1185}

\bibitem[\protect\citeauthoryear{{Libralato} et~al.,}{{Libralato}
  et~al.}{2016}]{Libralato2016}
{Libralato} M.,  et~al., 2016, \mn@doi [\mnras] {10.1093/mnras/stw1932}, \href
  {http://adsabs.harvard.edu/abs/2016MNRAS.463.1780L} {463, 1780}

\bibitem[\protect\citeauthoryear{{Livingston} et~al.,}{{Livingston}
  et~al.}{2018}]{Livingston2018}
{Livingston} J.~H.,  et~al., 2018, preprint, \href
  {http://adsabs.harvard.edu/abs/2018arXiv180901968L} {} (\mn@eprint {arXiv}
  {1809.01968})

\bibitem[\protect\citeauthoryear{{Lopez} \& {Fortney}}{{Lopez} \&
  {Fortney}}{2013}]{Lopez2013}
{Lopez} E.~D.,  {Fortney} J.~J.,  2013, \mn@doi [\apj]
  {10.1088/0004-637X/776/1/2}, \href
  {http://adsabs.harvard.edu/abs/2013ApJ...776....2L} {776, 2}

\bibitem[\protect\citeauthoryear{{Louden}, {Wheatley}  \& {Briggs}}{{Louden}
  et~al.}{2017}]{Louden2017}
{Louden} T.,  {Wheatley} P.~J.,   {Briggs} K.,  2017, \mn@doi [\mnras]
  {10.1093/mnras/stw2421}, \href
  {http://adsabs.harvard.edu/abs/2017MNRAS.464.2396L} {464, 2396}

\bibitem[\protect\citeauthoryear{{Luger}, {Agol}, {Kruse}, {Barnes}, {Becker},
  {Foreman-Mackey}  \& {Deming}}{{Luger} et~al.}{2016}]{Luger2016}
{Luger} R.,  {Agol} E.,  {Kruse} E.,  {Barnes} R.,  {Becker} A.,
  {Foreman-Mackey} D.,   {Deming} D.,  2016, \mn@doi [\aj]
  {10.3847/0004-6256/152/4/100}, \href
  {http://adsabs.harvard.edu/abs/2016AJ....152..100L} {152, 100}

\bibitem[\protect\citeauthoryear{{Luger}, {Kruse}, {Foreman-Mackey}, {Agol}  \&
  {Saunders}}{{Luger} et~al.}{2017}]{Luger2017}
{Luger} R.,  {Kruse} E.,  {Foreman-Mackey} D.,  {Agol} E.,   {Saunders} N.,
  2017, preprint, \href {http://adsabs.harvard.edu/abs/2017arXiv170205488L} {}
  (\mn@eprint {arXiv} {1702.05488})

\bibitem[\protect\citeauthoryear{{Lundkvist} et~al.,}{{Lundkvist}
  et~al.}{2016}]{Lundkvist2016}
{Lundkvist} M.~S.,  et~al., 2016, \mn@doi [Nature Communications]
  {10.1038/ncomms11201}, \href
  {http://adsabs.harvard.edu/abs/2016NatCo...711201L} {7, 11201}

\bibitem[\protect\citeauthoryear{{Mann} et~al.,}{{Mann}
  et~al.}{2017}]{Mann2017}
{Mann} A.~W.,  et~al., 2017, \mn@doi [\aj] {10.1088/1361-6528/aa5276}, \href
  {http://adsabs.harvard.edu/abs/2017AJ....153...64M} {153, 64}

\bibitem[\protect\citeauthoryear{{Martinez}, {Cunha}, {Ghezzi}  \&
  {Smith}}{{Martinez} et~al.}{2019}]{Martinez2019}
{Martinez} C.~F.,  {Cunha} K.,  {Ghezzi} L.,   {Smith} V.~V.,  2019, \mn@doi
  [\apj] {10.3847/1538-4357/ab0d93}, \href
  {https://ui.adsabs.harvard.edu/abs/2019ApJ...875...29M} {875, 29}

\bibitem[\protect\citeauthoryear{{Matsakos} \& {K{\"o}nigl}}{{Matsakos} \&
  {K{\"o}nigl}}{2016}]{Matsakos2016}
{Matsakos} T.,  {K{\"o}nigl} A.,  2016, \mn@doi [\apjl]
  {10.3847/2041-8205/820/1/L8}, \href
  {http://adsabs.harvard.edu/abs/2016ApJ...820L...8M} {820, L8}

\bibitem[\protect\citeauthoryear{{Mazeh}, {Holczer}  \& {Faigler}}{{Mazeh}
  et~al.}{2016}]{Mazeh2016}
{Mazeh} T.,  {Holczer} T.,   {Faigler} S.,  2016, \mn@doi [\aap]
  {10.1051/0004-6361/201528065}, \href
  {http://adsabs.harvard.edu/abs/2016A%26A...589A..75M} {589, A75}

\bibitem[\protect\citeauthoryear{{Micela}}{{Micela}}{2002}]{Micela2002}
{Micela} G.,  2002, in {Montesinos} B.,  {Gimenez} A.,   {Guinan} E.~F.,  eds,
  Astronomical Society of the Pacific Conference Series Vol. 269, The Evolving
  Sun and its Influence on Planetary Environments. p.~107

\bibitem[\protect\citeauthoryear{{Micela}, {Sciortino}, {Serio}, {Vaiana},
  {Bookbinder}, {Golub}, {Harnden}  \& {Rosner}}{{Micela}
  et~al.}{1985}]{Micela1985}
{Micela} G.,  {Sciortino} S.,  {Serio} S.,  {Vaiana} G.~S.,  {Bookbinder} J.,
  {Golub} L.,  {Harnden} Jr. F.~R.,   {Rosner} R.,  1985, \mn@doi [\apj]
  {10.1086/163143}, \href {http://adsabs.harvard.edu/abs/1985ApJ...292..172M}
  {292, 172}

\bibitem[\protect\citeauthoryear{{Newton} et~al.,}{{Newton}
  et~al.}{2019}]{Newton2019}
{Newton} E.~R.,  et~al., 2019, \mn@doi [\apjl] {10.3847/2041-8213/ab2988},
  \href {https://ui.adsabs.harvard.edu/abs/2019ApJ...880L..17N} {880, L17}

\bibitem[\protect\citeauthoryear{{Noyes}, {Weiss}  \& {Vaughan}}{{Noyes}
  et~al.}{1984}]{Noyes1984}
{Noyes} R.~W.,  {Weiss} N.~O.,   {Vaughan} A.~H.,  1984, \mn@doi [\apj]
  {10.1086/162735}, \href {http://adsabs.harvard.edu/abs/1984ApJ...287..769N}
  {287, 769}

\bibitem[\protect\citeauthoryear{{N{\'u}{\~n}ez} \&
  {Ag{\"u}eros}}{{N{\'u}{\~n}ez} \& {Ag{\"u}eros}}{2016}]{Nunez2016}
{N{\'u}{\~n}ez} A.,  {Ag{\"u}eros} M.~A.,  2016, \mn@doi [\apj]
  {10.3847/0004-637X/830/1/44}, \href
  {https://ui.adsabs.harvard.edu/abs/2016ApJ...830...44N} {830, 44}

\bibitem[\protect\citeauthoryear{{N{\'u}{\~n}ez} et~al.,}{{N{\'u}{\~n}ez}
  et~al.}{2015}]{Nunez2015}
{N{\'u}{\~n}ez} A.,  et~al., 2015, \mn@doi [\apj]
  {10.1088/0004-637X/809/2/161}, \href
  {https://ui.adsabs.harvard.edu/abs/2015ApJ...809..161N} {809, 161}

\bibitem[\protect\citeauthoryear{{Obermeier} et~al.,}{{Obermeier}
  et~al.}{2016}]{Obermeier2016}
{Obermeier} C.,  et~al., 2016, \mn@doi [\aj] {10.3847/1538-3881/152/6/223},
  \href {http://adsabs.harvard.edu/abs/2016AJ....152..223O} {152, 223}

\bibitem[\protect\citeauthoryear{{Owen} \& {Alvarez}}{{Owen} \&
  {Alvarez}}{2016}]{Owen2016}
{Owen} J.~E.,  {Alvarez} M.~A.,  2016, \mn@doi [\apj]
  {10.3847/0004-637X/816/1/34}, \href
  {https://ui.adsabs.harvard.edu/abs/2016ApJ...816...34O} {816, 34}

\bibitem[\protect\citeauthoryear{{Owen} \& {Lai}}{{Owen} \&
  {Lai}}{2018}]{Owen2018}
{Owen} J.~E.,  {Lai} D.,  2018, \mn@doi [\mnras] {10.1093/mnras/sty1760}, \href
  {http://adsabs.harvard.edu/abs/2018MNRAS.tmp.1685O} {}

\bibitem[\protect\citeauthoryear{{Owen} \& {Wu}}{{Owen} \&
  {Wu}}{2013}]{Owen2013}
{Owen} J.~E.,  {Wu} Y.,  2013, \mn@doi [\apj] {10.1088/0004-637X/775/2/105},
  \href {http://adsabs.harvard.edu/abs/2013ApJ...775..105O} {775, 105}

\bibitem[\protect\citeauthoryear{{Owen} \& {Wu}}{{Owen} \&
  {Wu}}{2017}]{Owen2017}
{Owen} J.~E.,  {Wu} Y.,  2017, \mn@doi [\apj] {10.3847/1538-4357/aa890a}, \href
  {http://adsabs.harvard.edu/abs/2017ApJ...847...29O} {847, 29}

\bibitem[\protect\citeauthoryear{{Pallavicini}, {Golub}, {Rosner}, {Vaiana},
  {Ayres}  \& {Linsky}}{{Pallavicini} et~al.}{1981}]{Pallavicini1981}
{Pallavicini} R.,  {Golub} L.,  {Rosner} R.,  {Vaiana} G.~S.,  {Ayres} T.,
  {Linsky} J.~L.,  1981, \mn@doi [\apj] {10.1086/159152}, \href
  {http://adsabs.harvard.edu/abs/1981ApJ...248..279P} {248, 279}

\bibitem[\protect\citeauthoryear{{Pizzolato}, {Maggio}, {Micela}, {Sciortino}
  \& {Ventura}}{{Pizzolato} et~al.}{2003}]{Pizzolato2003}
{Pizzolato} N.,  {Maggio} A.,  {Micela} G.,  {Sciortino} S.,   {Ventura} P.,
  2003, \mn@doi [\aap] {10.1051/0004-6361:20021560}, \href
  {http://adsabs.harvard.edu/abs/2003A%26A...397..147P} {397, 147}

\bibitem[\protect\citeauthoryear{{Pope}, {Parviainen}  \& {Aigrain}}{{Pope}
  et~al.}{2016}]{Pope2016}
{Pope} B.~J.~S.,  {Parviainen} H.,   {Aigrain} S.,  2016, \mn@doi [\mnras]
  {10.1093/mnras/stw1373}, \href
  {http://adsabs.harvard.edu/abs/2016MNRAS.461.3399P} {461, 3399}

\bibitem[\protect\citeauthoryear{{Rizzuto}, {Vanderburg}, {Mann}, {Kraus},
  {Dressing}, {Ag{\"u}eros}, {Douglas}  \& {Krolikowski}}{{Rizzuto}
  et~al.}{2018}]{Rizzuto2018}
{Rizzuto} A.~C.,  {Vanderburg} A.,  {Mann} A.~W.,  {Kraus} A.~L.,  {Dressing}
  C.~D.,  {Ag{\"u}eros} M.~A.,  {Douglas} S.~T.,   {Krolikowski} D.~M.,  2018,
  \mn@doi [\aj] {10.3847/1538-3881/aadf37}, \href
  {https://ui.adsabs.harvard.edu/abs/2018AJ....156..195R} {156, 195}

\bibitem[\protect\citeauthoryear{{Rogers}, {Gupta}, {Owen}  \&
  {Schlichting}}{{Rogers} et~al.}{2021}]{Rogers2021}
{Rogers} J.~G.,  {Gupta} A.,  {Owen} J.~E.,   {Schlichting} H.~E.,  2021, arXiv
  e-prints, \href {https://ui.adsabs.harvard.edu/abs/2021arXiv210503443R} {p.
  arXiv:2105.03443}

\bibitem[\protect\citeauthoryear{{Salz}, {Schneider}, {Czesla}  \&
  {Schmitt}}{{Salz} et~al.}{2015}]{Salz2015}
{Salz} M.,  {Schneider} P.~C.,  {Czesla} S.,   {Schmitt} J.~H.~M.~M.,  2015,
  \mn@doi [\aap] {10.1051/0004-6361/201425243}, \href
  {http://adsabs.harvard.edu/abs/2015A%26A...576A..42S} {576, A42}

\bibitem[\protect\citeauthoryear{{Sanz-Forcada}, {Micela}, {Ribas}, {Pollock},
  {Eiroa}, {Velasco}, {Solano}  \& {Garc{\'{\i}}a-{\'A}lvarez}}{{Sanz-Forcada}
  et~al.}{2011}]{SanzForcada2011}
{Sanz-Forcada} J.,  {Micela} G.,  {Ribas} I.,  {Pollock} A.~M.~T.,  {Eiroa} C.,
   {Velasco} A.,  {Solano} E.,   {Garc{\'{\i}}a-{\'A}lvarez} D.,  2011, \mn@doi
  [\aap] {10.1051/0004-6361/201116594}, \href
  {http://adsabs.harvard.edu/abs/2011A%26A...532A...6S} {532, A6}

\bibitem[\protect\citeauthoryear{{Schultz} \& {Wiemer}}{{Schultz} \&
  {Wiemer}}{1975}]{Schultz1975}
{Schultz} G.~V.,  {Wiemer} W.,  1975, \aap, \href
  {https://ui.adsabs.harvard.edu/abs/1975A&A....43..133S} {43, 133}

\bibitem[\protect\citeauthoryear{{Smith}, {Brickhouse}, {Liedahl}  \&
  {Raymond}}{{Smith} et~al.}{2001}]{Smith2001}
{Smith} R.~K.,  {Brickhouse} N.~S.,  {Liedahl} D.~A.,   {Raymond} J.~C.,  2001,
  \mn@doi [\apjl] {10.1086/322992}, \href
  {http://adsabs.harvard.edu/abs/2001ApJ...556L..91S} {556, L91}

\bibitem[\protect\citeauthoryear{{Soderblom}}{{Soderblom}}{2010}]{Soderblom2010}
{Soderblom} D.~R.,  2010, \mn@doi [\araa]
  {10.1146/annurev-astro-081309-130806}, \href
  {http://adsabs.harvard.edu/abs/2010ARA%26A..48..581S} {48, 581}

\bibitem[\protect\citeauthoryear{{Szab{\'o}} \& {Kiss}}{{Szab{\'o}} \&
  {Kiss}}{2011}]{Szabo2011}
{Szab{\'o}} G.~M.,  {Kiss} L.~L.,  2011, \mn@doi [\apjl]
  {10.1088/2041-8205/727/2/L44}, \href
  {http://adsabs.harvard.edu/abs/2011ApJ...727L..44S} {727, L44}

\bibitem[\protect\citeauthoryear{{Taylor}}{{Taylor}}{2006}]{Taylor2006}
{Taylor} B.~J.,  2006, \mn@doi [\aj] {10.1086/508610}, \href
  {https://ui.adsabs.harvard.edu/abs/2006AJ....132.2453T} {132, 2453}

\bibitem[\protect\citeauthoryear{{Tofflemire} et~al.,}{{Tofflemire}
  et~al.}{2021}]{Tofflemire2021}
{Tofflemire} B.~M.,  et~al., 2021, \mn@doi [\aj] {10.3847/1538-3881/abdf53},
  \href {https://ui.adsabs.harvard.edu/abs/2021AJ....161..171T} {161, 171}

\bibitem[\protect\citeauthoryear{{Van Eylen}, {Agentoft}, {Lundkvist},
  {Kjeldsen}, {Owen}, {Fulton}, {Petigura}  \& {Snellen}}{{Van Eylen}
  et~al.}{2018}]{VanEylen2018}
{Van Eylen} V.,  {Agentoft} C.,  {Lundkvist} M.~S.,  {Kjeldsen} H.,  {Owen}
  J.~E.,  {Fulton} B.~J.,  {Petigura} E.,   {Snellen} I.,  2018, \mn@doi
  [\mnras] {10.1093/mnras/sty1783}, \href
  {https://ui.adsabs.harvard.edu/#abs/2018MNRAS.tmp.1712V} {p.~1712}

\bibitem[\protect\citeauthoryear{{Venturini}, {Guilera}, {Haldemann}, {Ronco}
  \& {Mordasini}}{{Venturini} et~al.}{2020}]{Venturini2020}
{Venturini} J.,  {Guilera} O.~M.,  {Haldemann} J.,  {Ronco} M.~P.,
  {Mordasini} C.,  2020, \mn@doi [\aap] {10.1051/0004-6361/202039141}, \href
  {https://ui.adsabs.harvard.edu/abs/2020A&A...643L...1V} {643, L1}

\bibitem[\protect\citeauthoryear{{Walsh}, {Kuntz}, {Collier}, {Sibeck},
  {Snowden}  \& {Thomas}}{{Walsh} et~al.}{2014}]{Walsh2014}
{Walsh} B.~M.,  {Kuntz} K.~D.,  {Collier} M.~R.,  {Sibeck} D.~G.,  {Snowden}
  S.~L.,   {Thomas} N.~E.,  2014, \mn@doi [Space Weather]
  {10.1002/2014SW001046}, \href
  {http://adsabs.harvard.edu/abs/2014SpWea..12..387W} {12, 387}

\bibitem[\protect\citeauthoryear{{Wang} et~al.,}{{Wang}
  et~al.}{2014}]{Wang2014}
{Wang} P.~F.,  et~al., 2014, \mn@doi [\apj] {10.1088/0004-637X/784/1/57}, \href
  {https://ui.adsabs.harvard.edu/abs/2014ApJ...784...57W} {784, 57}

\bibitem[\protect\citeauthoryear{{Watson}, {Donahue}  \& {Walker}}{{Watson}
  et~al.}{1981}]{Watson1981}
{Watson} A.~J.,  {Donahue} T.~M.,   {Walker} J.~C.~G.,  1981, \mn@doi [\icarus]
  {10.1016/0019-1035(81)90101-9}, \href
  {https://ui.adsabs.harvard.edu/abs/1981Icar...48..150W} {48, 150}

\bibitem[\protect\citeauthoryear{{Wheatley}, {Louden}, {Bourrier}, {Ehrenreich}
   \& {Gillon}}{{Wheatley} et~al.}{2017}]{Wheatley2017}
{Wheatley} P.~J.,  {Louden} T.,  {Bourrier} V.,  {Ehrenreich} D.,   {Gillon}
  M.,  2017, \mn@doi [\mnras] {10.1093/mnrasl/slw192}, \href
  {http://adsabs.harvard.edu/abs/2017MNRAS.465L..74W} {465, L74}

\bibitem[\protect\citeauthoryear{{Wilms}, {Allen}  \& {McCray}}{{Wilms}
  et~al.}{2000}]{Wilms2000}
{Wilms} J.,  {Allen} A.,   {McCray} R.,  2000, \mn@doi [\apj] {10.1086/317016},
  \href {http://adsabs.harvard.edu/abs/2000ApJ...542..914W} {542, 914}

\bibitem[\protect\citeauthoryear{{Wolfgang}, {Rogers}  \& {Ford}}{{Wolfgang}
  et~al.}{2016}]{Wolfgang2016}
{Wolfgang} A.,  {Rogers} L.~A.,   {Ford} E.~B.,  2016, \mn@doi [\apj]
  {10.3847/0004-637X/825/1/19}, \href
  {http://adsabs.harvard.edu/abs/2016ApJ...825...19W} {825, 19}

\bibitem[\protect\citeauthoryear{{Wright} \& {Drake}}{{Wright} \&
  {Drake}}{2016}]{Wright2016}
{Wright} N.~J.,  {Drake} J.~J.,  2016, \mn@doi [\nat] {10.1038/nature18638},
  \href {http://adsabs.harvard.edu/abs/2016Natur.535..526W} {535, 526}

\bibitem[\protect\citeauthoryear{{Wright}, {Drake}, {Mamajek}  \&
  {Henry}}{{Wright} et~al.}{2011}]{Wright2011}
{Wright} N.~J.,  {Drake} J.~J.,  {Mamajek} E.~E.,   {Henry} G.~W.,  2011,
  \mn@doi [\apj] {10.1088/0004-637X/743/1/48}, \href
  {http://adsabs.harvard.edu/abs/2011ApJ...743...48W} {743, 48}

\bibitem[\protect\citeauthoryear{{Wright}, {Newton}, {Williams}, {Drake}  \&
  {Yadav}}{{Wright} et~al.}{2018}]{Wright2018}
{Wright} N.~J.,  {Newton} E.~R.,  {Williams} P. K.~G.,  {Drake} J.~J.,
  {Yadav} R.~K.,  2018, \mn@doi [\mnras] {10.1093/mnras/sty1670}, \href
  {https://ui.adsabs.harvard.edu/abs/2018MNRAS.479.2351W} {479, 2351}

\makeatother
\end{thebibliography}








\bsp	
\label{lastpage}
\end{document}